\documentclass{aa}

\usepackage{graphicx}
\graphicspath{{figs/}}
\usepackage[separate-uncertainty]{siunitx}
\usepackage{txfonts}
\usepackage{xcolor}
\usepackage{ulem}
\usepackage{hyperref}
\usepackage[all]{hypcap}
\usepackage[capitalise]{cleveref}
\usepackage{placeins}
\usepackage{stfloats}
\usepackage{lineno}
\nolinenumbers

\begin{document}

   \title{The SRG/eROSITA All-Sky Survey}
   \subtitle{View of the Fornax galaxy cluster}
   \titlerunning{eROSITA view of the Fornax cluster}

   \author{T.H. Reiprich\inst{1}
          \and A. Veronica\inst{1}
          \and F. Pacaud\inst{1}
          \and P. St\"ocker\inst{1}
          \and V. Nazaretyan\inst{1}
          \and A. Srivastava\inst{1}
          \and A. Pandya\inst{1}
          \and J. Dietl\inst{1}
          \and J.S. Sanders\inst{2}
          \and M.C.H. Yeung\inst{2}
          \and A. Chaturvedi\inst{3}
          \and M. Hilker\inst{4}
          \and B. Seidel\inst{5}
          \and K. Dolag\inst{5}
          \and J. Comparat\inst{2}
          \and V. Ghirardini\inst{6,2}
          \and M. Kluge\inst{2}
          \and A. Liu\inst{7,2}
          \and N. Malavasi\inst{2}
          \and X. Zhang\inst{2}
          \and E. Hern{\'a}ndez-Mart{\'i}nez\inst{5}
          }
    \authorrunning{T.H. Reiprich et al.}

   \institute{Argelander-Institut f\"ur Astronomie (AIfA), Universit\"at Bonn, Auf dem H\"ugel 71, 53121 Bonn, Germany
   \and
   Max-Planck-Institut f\"ur extraterrestrische Physik, Gie{\ss}enbachstr.~1, 85748 Garching bei M\"unchen, Germany
   \and
   Leibniz-Institut f\"ur Astrophysik Potsdam (AIP), An der Sternwarte 16, D-14482 Potsdam, Germany
   \and
   European Southern Observatory, Karl-Schwarzschild-Stra{\ss}e 2, 85748 Garching, Germany
   \and
   Universit\"ats-Sternwarte, Fakult\"at für Physik, Ludwig-Maximilians-Universit\"at M\"unchen, Scheinerstr.~1, 81679 M\"unchen, Germany
   \and 
   INAF, Osservatorio di Astrofisica e Scienza dello Spazio, via Piero Gobetti 93/3, 40129 Bologna, Italy
   \and
   Institute for Frontiers in Astronomy and Astrophysics, Beijing Normal University, Beijing 102206, China
}
   \date{Received ...; accepted ...}
 
  \abstract
   {The Fornax cluster is one of the closest X-ray-bright galaxy clusters; as such, we can study the system at high spatial resolution. However, previous observations of the intracluster medium were limited to less than $R_{500}$.} 
   {We aim to significantly extend the X-ray coverage of the Fornax cluster and to search for features in the X-ray surface brightness distribution beyond $R_{500}$ induced by the gravitational growth of this system.}
   {We used data from five SRG/eROSITA all-sky surveys and performed a detailed one- and two-dimensional X-ray surface brightness analysis, tracing hot gas emission from kiloparsec to megaparsec scales with a single instrument. We compared the results to those from a recent numerical simulation of the local Universe (SLOW) and correlated the X-ray emission distribution with that of other tracers, including cluster member galaxies, ultra-compact dwarf galaxies, intracluster globular clusters, and HI-tail galaxies.}
   {We detect X-ray emission out to well beyond the virial radius, $R_{100}=2.2$ deg.
   In the inner regions within $R_{500}$, we see previously known features, such as a large-scale spiral-shaped edge; however, we do not find obvious evidence of the bow shock several hundred kiloparsecs south of the cluster center predicted by previous numerical simulations of the Fornax cluster.
   Instead, we discover emission fingers beyond $R_{500}$ to the west and southeast
   and excesses that stretch out far beyond the virial radius.
   They might be due to gas being pushed outward by the previous merger with NGC\,1404 or due to warm-hot gas infall along large-scale filaments. Intriguingly, we find the distributions of the other tracers -- galaxies and globular clusters -- to be correlated with the X-ray-excess regions, favoring the infall scenario. Interestingly, we also discover an apparent bridge of low-surface-brightness emission beyond the virial radius connecting to the Fornax\,A galaxy group, which is also traced by the member galaxy and globular cluster distribution. This X-ray bridge furthermore approximately coincides with a region of enhanced Faraday depth detected previously. The gas distribution in the SLOW simulation shows similar features as those we have discovered with SRG/eROSITA.
}
   {SRG/eROSITA has enabled us to tremendously expand the view of the intracluster medium of the Fornax cluster. We witness the growth of a cluster along large-scale filaments.}


   \maketitle
%

\section{Introduction}
\label{sec:intro}

Massive galaxy clusters are the youngest objects in the Universe. They continue to form and grow even today by merging with other clusters and via continuous accretion along large-scale filaments connecting clusters. Cluster outskirts around the virial radius trace this growth and the filament connection \citep[e.g.,][]{bbr11,rbe13, wsn19}. The intergalactic medium in clusters -- the intracluster medium (ICM) -- and in filaments -- the warm-hot intergalactic medium (WHIM) -- reaches X-ray emitting temperatures above $10^6$ K \citep[e.g.,][]{2019MNRAS.486.3766M}. The ICM in the inner parts of clusters has been studied for several decades, while their outskirts became accessible less than 20 years ago (see the review articles above and, e.g., the recent studies of the Virgo and Centaurus clusters, \citealt{mrv24,vrp24}). The
X-ray emission from the WHIM has been discovered only recently. For instance, WHIM filament discoveries around individual nearby cluster systems have been reported by \citet{rvp21} and \citet{dpr24} and through statistical stacking analyses by, for example, \citet{2024A&A...691A.234Z}. The nearby Fornax cluster is therefore an obvious candidate for a cluster outskirts study and filament search.

With a virial diameter of almost 4.5 deg, the Fornax cluster is one of the most extended clusters of galaxies in the sky. In fact, only the Virgo cluster \citep[e.g.,][]{mrv24} spans a larger area. The apparent large extent is due to their proximity. Their intrinsic physical sizes, i.e., masses, are actually not extraordinary; in fact, the Fornax cluster should rather be called a group of galaxies based on its mass of $<$$10^{14}\,\mathrm{M_\odot}$ \citep[e.g.,][]{1997ApJ...482..143J,2001ApJ...548L.139D,rb01,pap11,2024MNRAS.530.3787S}.
Given its proximity, the Fornax cluster can be studied at high spatial resolution and down to low surface brightnesses at all wavelengths. And, indeed, it is one of the best studied galaxy clusters in existence. In particular, its two central bright elliptical galaxies, NGC\,1399 and NGC\,1404, as well as the central galaxy of the neighboring Fornax\,A galaxy group, NGC\,1316, have been the subject of many previous studies.

For example,
\citet{2017ApJ...851...69S} used a mosaic of 15 \textit{XMM-Newton} observations (see also \citealt{2011PASJ...63S.963M}) and \textit{Chandra} observations to describe four contact discontinuities identified as sloshing cold fronts \citep[for a review of cold fronts, see, e.g.,][]{mv07}. They suggest these cold fronts are caused by the infall of NGC\,1404.
Deep \textit{Chandra} data have been used to study NGC\,1404 at high spatial resolution, resulting in the findings of a sharp contact discontinuity to the northwest, i.e., in the direction of NGC\,1399 at the center of the Fornax cluster, and a tail to the southeast \citep[e.g.,][]{2005ApJ...621..663M,2017ApJ...834...74S}, consistent with NGC\,1404 falling into the core of the Fornax cluster as already indicated by ROSAT PSPC data \citep{1997ApJ...482..143J}.

\citet{2018ApJ...865..118S} then used detailed hydrodynamic simulations to determine whether a presumed infall of NGC\,1404 could give rise to the central ICM morphology of the Fornax cluster, for example the cold fronts described above as well as the contact discontinuity and tail of NGC\,1404. And, indeed, they found the most likely scenario to be that NGC\,1404 is on its second or third encounter with the Fornax center. Furthermore, they predicted a detached bow shock 450 kpc to 750 kpc south of NGC\,1404. With the eROSITA data used here, we put this prediction to the test.

NGC\,1316 is the central galaxy of a group of galaxies that, based on galaxy dynamics, appears to be bound to the Fornax cluster and infalling onto it for the first time  \citep{2001ApJ...548L.139D}.
This group is dominated by late-type galaxies, and its low fraction of intragroup light points to an ongoing phase of preprocessing \citep{2020A&A...640A.137R}.
NGC\,1316 also hosts the well-known Fornax\,A extended radio source, and its radio morphology is consistent with a northward, i.e., infalling, motion \citep[e.g.,][]{1983A&A...127..361E}.
Recently, \citet{2021PASA...38...20A} studied the Faraday depth in a large field surrounding the Fornax cluster using a large number of background sources (see also \citealt{2025A&A...694A.125L} for a smaller but more densely sampled Faraday rotation measure grid). \citeauthor{2021PASA...38...20A} found a region of increased Faraday depth northeast of NGC\,1316, i.e., in the direction of NGC\,1399.
In light of these findings, it is of interest to search for warm-hot gas emission beyond the virial radius of the Fornax cluster in the direction of Fornax\,A, which is what we do with the eROSITA data analyzed here.

Given the higher velocity dispersion compared to big galaxies, \citet{2001ApJ...548L.139D} concluded that the dwarf galaxy population includes many infalling objects.
The spatial distribution of dwarf galaxies in the main Fornax cluster is highly structured along clumpy overdensities \citep[e.g.,][]{2018ApJ...859...52O}. The distribution of dwarf galaxies in the infalling Fornax\,A group is lopsided toward the main cluster, further evidence of a recent interaction between the two components \citep{2021A&A...647A.100S}. Given these findings, it is of interest to compare the spatial distribution of dwarf galaxies
with regions of excess X-ray emission and the warm-hot gas distribution in the cluster outskirts, which may indicate preferred filamentary inflow directions. 

We also compared these distributions to the globular cluster and normal galaxy distributions.
The inhomogeneous and partly patchy spatial distributions of the intracluster light \citep{2016ApJ...820...42I, 2017ApJ...851...75I} and globular clusters \citep{2016ApJ...819L..31D, 2020A&A...639A.136C} within the virial radius trace the recent assembly history of the Fornax cluster. Kinematically cold substructures of intracluster globular clusters \citep{2008A&A...477L...9S, 2022A&A...657A..93C, 2022A&A...657A..94N} are evidence in favor of the assembly of the cluster along filaments.

The X-ray telescope eROSITA \citep[][]{paa21} on board the Spectrum-Roentgen-Gamma (SRG) satellite \citep{2021A&A...656A.132S} has performed four all-sky surveys to date, plus one partial sky survey \citep{2022SPIE12181E..1AC}. The data from the first all-sky survey and the corresponding X-ray source catalog are published \citep[][]{Merloni_2024}. Given the unlimited field of view, these data are perfectly suited to studying the very extended Fornax cluster and its surroundings. In this paper, we use data from five surveys.
The four additional surveys are each of very similar quality as the first survey. Some minor changes include a decreased particle-induced background \citep[PIB;][]{2023A&A...676A...3Y}, an increased frequency of flares, and some additional bad pixels for telescope module (TM) 4 due to micrometeoroid impact \citep{2022SPIE12181E..55F}.

\begin{figure}
    \centering
    \includegraphics[width=\hsize]{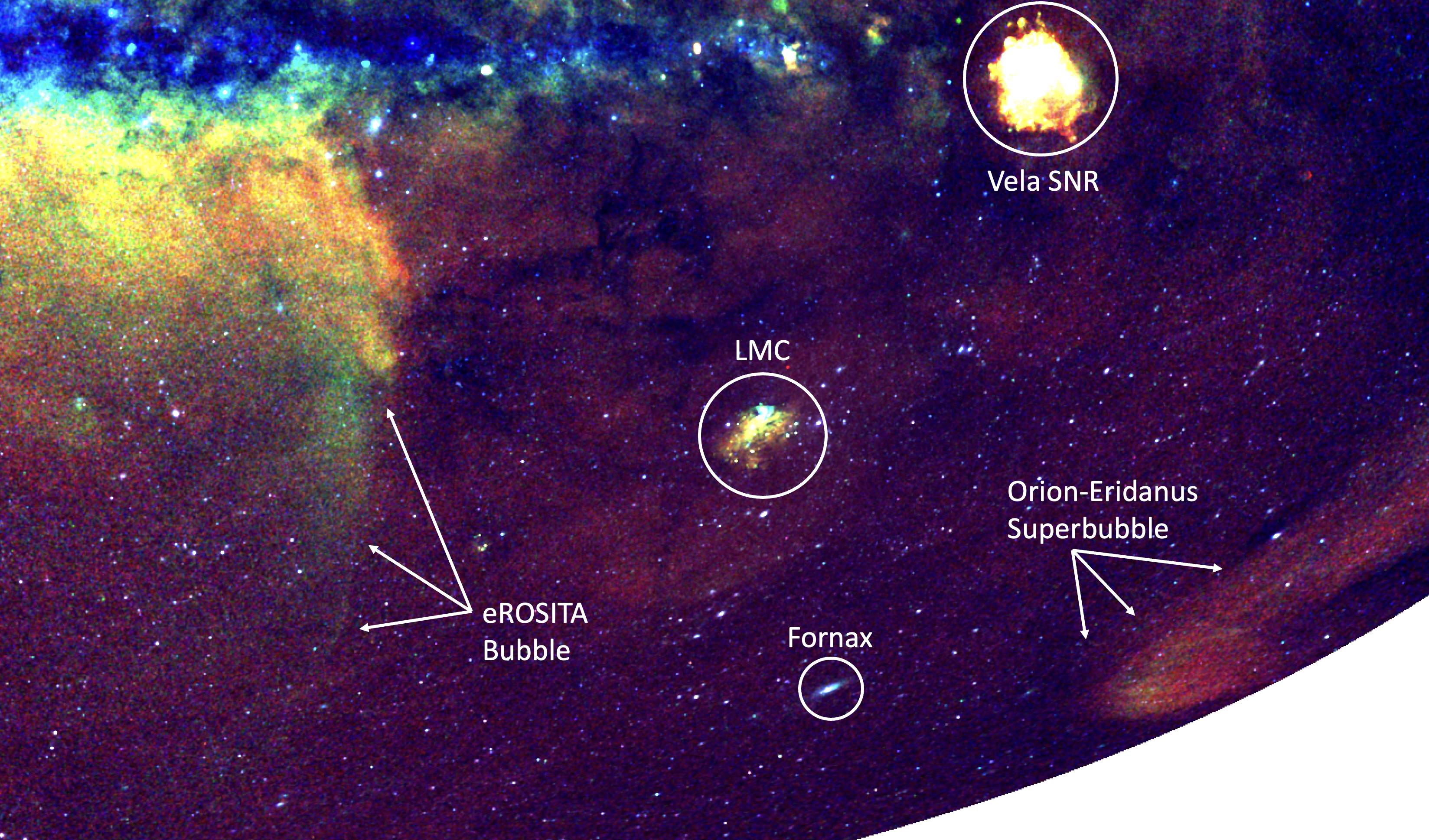}
    \caption{eRASS1 broadband maps in Galactic coordinates from \citet{2024A&A...681A..77Z} in the energy bands 0.4--0.6 keV (red), 0.6--1.0 keV (green), and 1.0--2.3 keV (blue). Some prominent extended sources are labeled, including the Large Magellanic Cloud (LMC). The nearby bright Fornax cluster lies in a particularly benign X-ray background region, optimally placed for studies of its low surface brightness outskirts.}
    \label{fig:eRASS1_map}
\end{figure}

According to the MCXC database \citep{pap11}, the Fornax cluster is located at RA\ $=$ 54.6163 and Dec.\ $=-$35.4483 (J2000) at a redshift of 0.005 and has $R_{500}=1.06$ deg $= 393$ kpc. $R_{500}$ is the radius within which the mean total density equals 500 times the critical density.
Using the conversion in \citet{rbe13}, this implies $R_{200}=1.63$ deg $=604$ kpc
and $R_{100}=2.22$ deg $=816$ kpc;
i.e., the virial region of the Fornax cluster spans almost 4.5 deg across. The Fornax cluster lies projected in a region of the sky that features a rather simple X-ray background structure (Fig.~\ref{fig:eRASS1_map})\footnote{\url{https://erosita.mpe.mpg.de/dr1/AllSkySurveyData_dr1/HalfSkyMaps_dr1/}}, enabling us to perform a robust background treatment despite the large extent.

Throughout this paper we assume a flat cosmology with $\Omega_{\rm m}=0.3$ and $H_0=70$ km/s/Mpc.\ For a redshift of 0.005, 1 arcsec corresponds to 0.103 kpc.

\section{eROSITA data reduction and analysis}
\label{sec:red}
Since we were interested in the faintest emission regions of the cluster outskirts, we took advantage of eROSITA data from all five currently existing sky surveys, based on processing version c020. For the data reduction, we used the eROSITA Science Analysis Software System \citep[eSASS,][]{2022A&A...661A...1B} version users\_211214.0.4.

We closely  followed the steps described in detail in \citet{rvp21}, \citet{2024A&A...681A.108V,vrp24}, \citet{mrv24}, and \citet{dpr24}. Therefore, here we only briefly outline these steps.

We employed a sufficient number of tiles around the Fornax cluster center to cover an area at least out to $\sim$$2R_{100}$ in all directions.
We used data from all seven TMs and applied the parameters \texttt{pattern=15} and \texttt{flag=0xe00fff30} when running \texttt{evtool}. We used the latter strict flag filtering since we were mostly interested in the faintest regions and therefore required the cleanest data.\footnote{See \url{https://erosita.mpe.mpg.de/edr/DataAnalysis/prod_descript/EventFiles_edr.html} for details on flag settings.} We ran \texttt{flaregti} using photons with energy above 5 keV and setting a $3\sigma$ threshold for the light curve filtering. Images and vignetted exposure maps were then created in the energy band 0.2--2.3~keV (0.8--2.3~keV) for the TMs with (without) on-chip filter \citep{paa21}. The PIB was assumed constant across the detector \citep{2020SPIE11444E..1OF}, and we modeled the PIB map of each TM as a rescaled version of its un-vignetted exposure map. The PIB normalizations were estimated from the observed hard band 6.0--9.0~keV counts, rescaled to the appropriate energy range using the spectral distribution of PIB measured in the filter-wheel-closed data \citep{2023A&A...676A...3Y}. To generate false color (RGB) images, images and exposure maps in additional bands were also created. The final images were fully corrected: (i) the PIB was subtracted; (ii) a relative correction for the varying foreground absorption in the field was performed using data from \citet{2016A&A...594A.116H}; and (iii) the different TM sensitivities and energy bands were taken into account for the total exposure correction, including vignetting.

\begin{figure}
    \centering
    \includegraphics[width=\hsize]{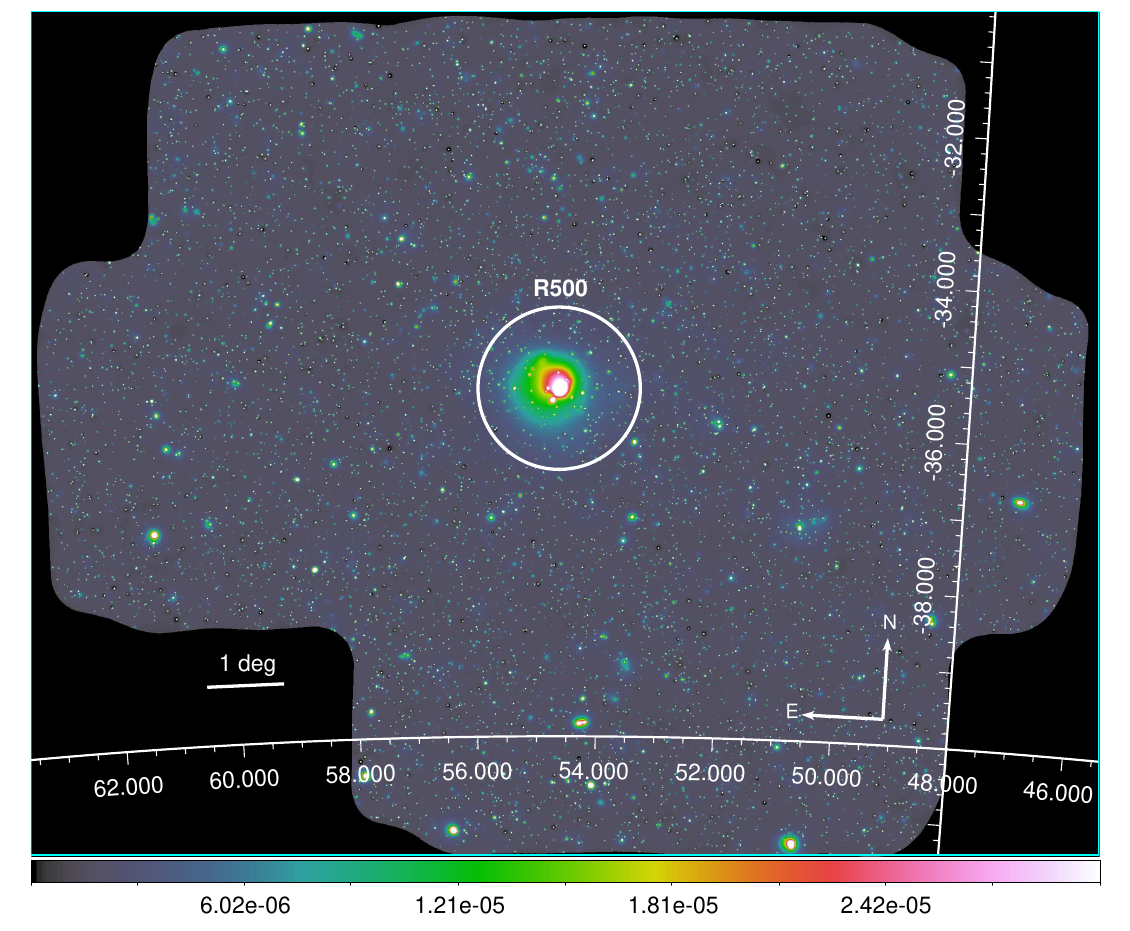}
    \includegraphics[width=\hsize]{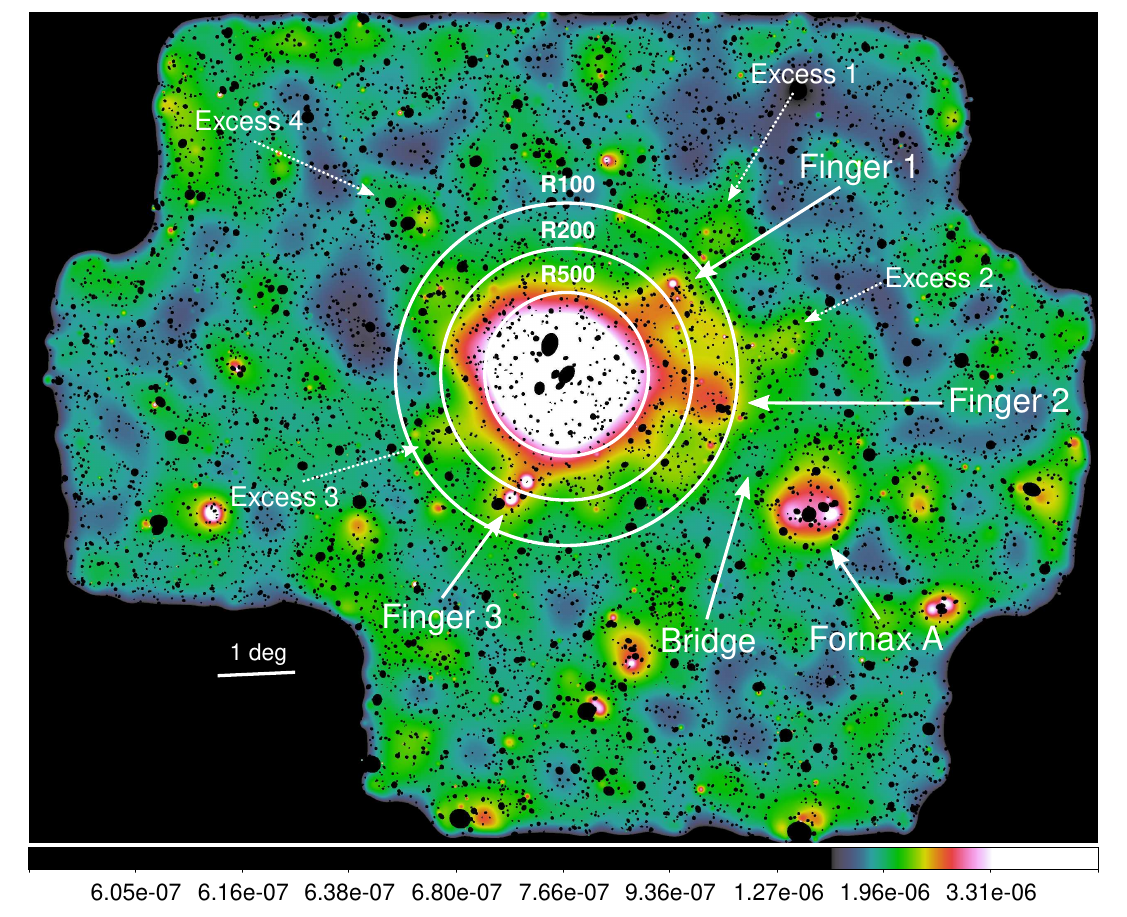}
    \caption{Fully corrected, wavelet-filtered image in the energy band 0.2--2.3 keV; characteristic radii are overlaid. Top: Linear scale to enhance the central surface brightness structure. Bottom: Log scale to enhance the outer low surface brightness structure; sources of small extents were excised. The emission features described in Sect.~\ref{sec:res} are labeled.}
    \label{fig:image}
\end{figure}

Since we were primarily interested in very low surface brightness regions, we performed wavelet-filtering to reduce the noise in the final images as described in \citet{rvp21}, using an implementation by F.\ Pacaud of the filtering method in the \texttt{MR/1} software \citep{1998A&AS..128..397S}.
For some images, we excised sources detected with SExtractor \citep{1996A&AS..117..393B} on the wavelet-filtered images, exploiting the elliptical shape information. Figure~\ref{fig:image} shows two image versions resulting from this process (the labeled features are described in detail in Sect.~\ref{sec:res}).

\begin{figure}
    \centering
    \includegraphics[width=\hsize]{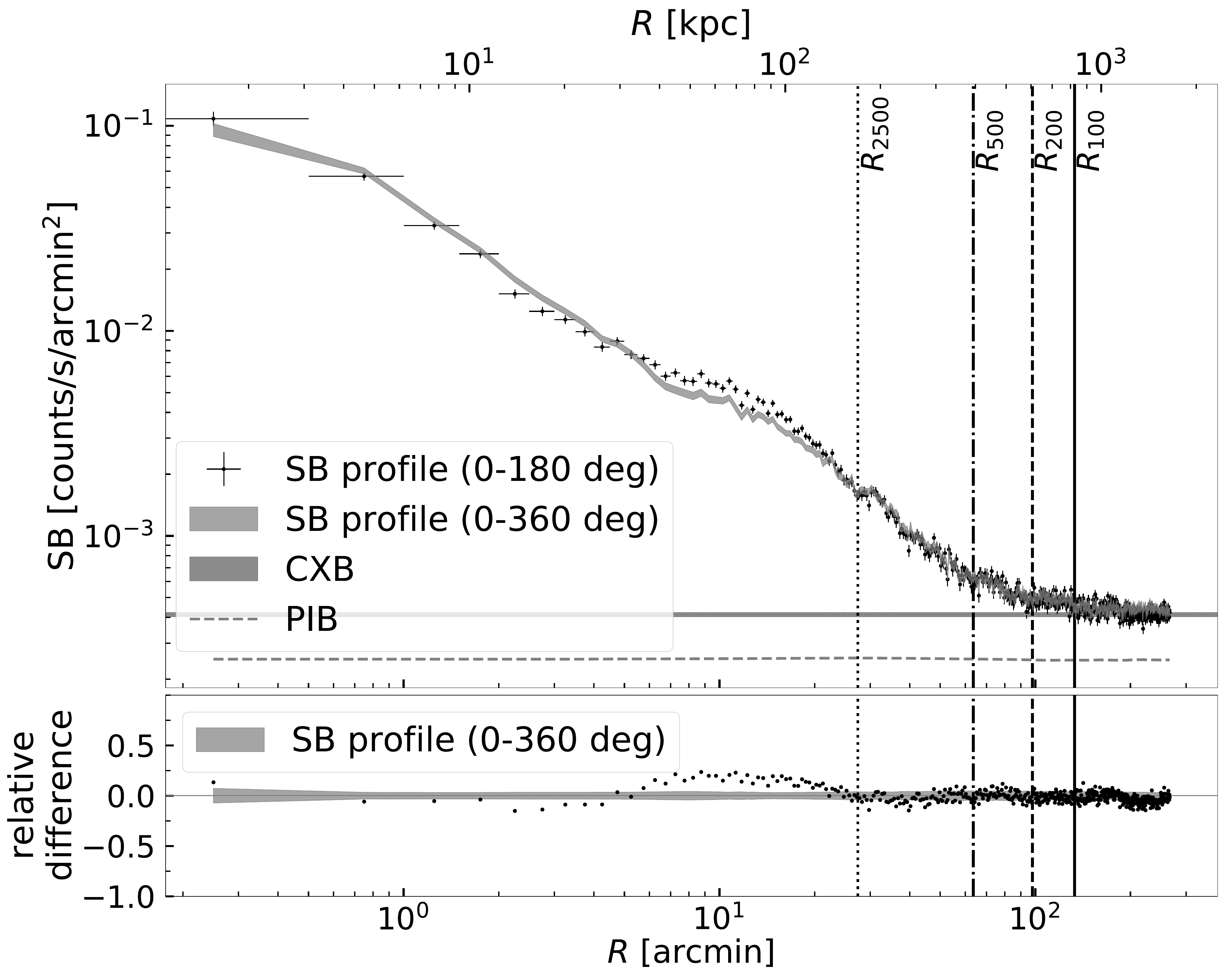}
    \includegraphics[width=\hsize]{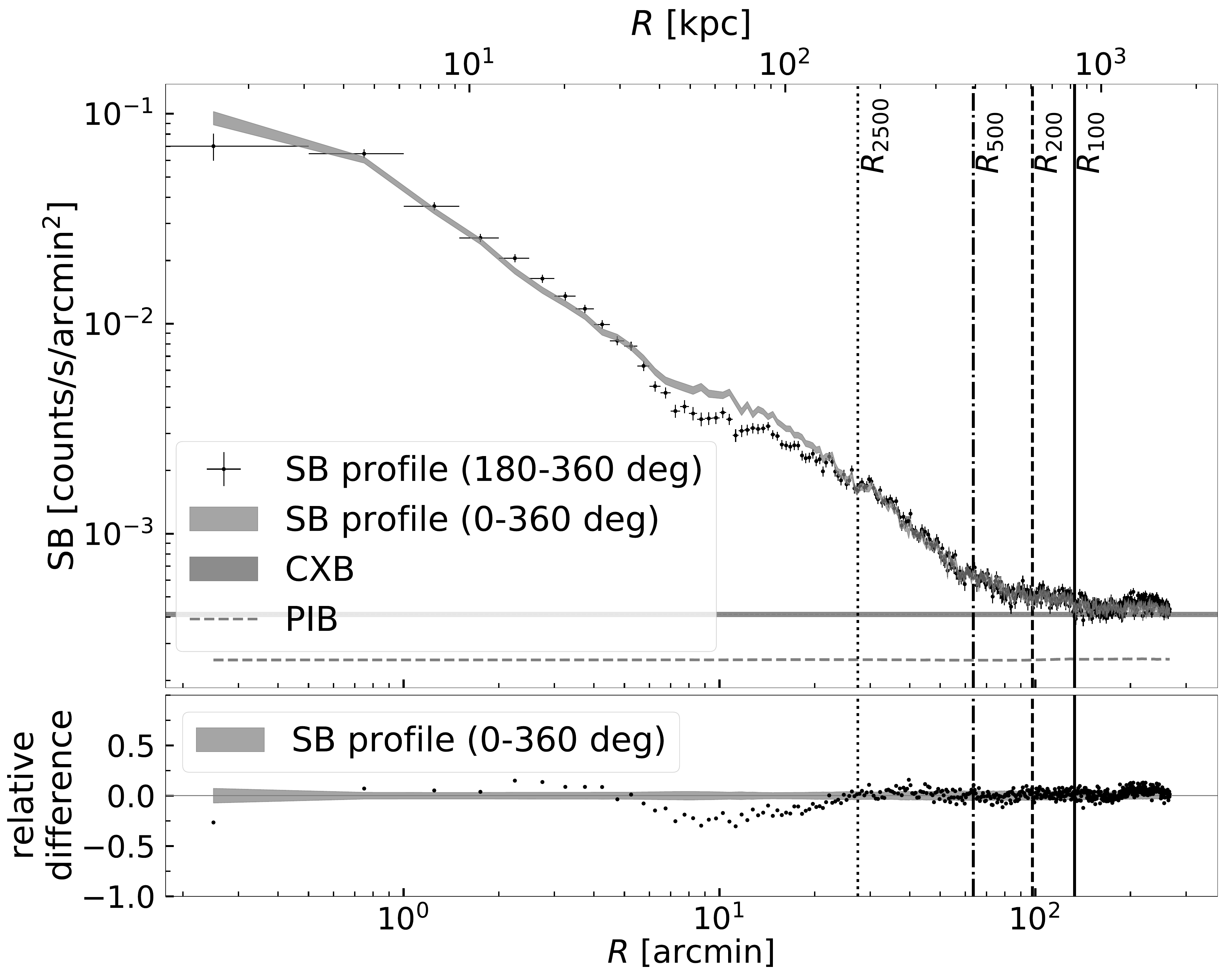}
    \caption{Fully corrected surface brightness profiles for two 180 deg wide sectors in comparison to that from the full (0--360 deg) annuli. The PIB is subtracted but not the CXB. Characteristic radii are overlaid. 0 deg is on the right (3 pm), 90 deg on the top (12 pm), 180 deg on the left (9 pm), and 270 deg on the bottom (6 pm).}
    \label{fig:SB_prof1}
\end{figure}

In addition to qualitative imaging analyses, we performed quantitative X-ray surface brightness analyses. We generated exposure- and relative absorption-corrected, PIB-subtracted, and compact source-excised surface brightness profiles in various sectors out to $2R_{100}$. For the profile analysis, we decreased the region of the SExtractor-detected source at the center of the Fornax cluster (NGC\,1399) to a circle with radius 30$''$ to remove mostly any potential emission from an active nucleus.
The level and uncertainty of the cosmic X-ray background (CXB) was determined by placing 11 boxes of size 1 deg$^2$ outside $\sim$$2R_{100}$ and calculating the average surface brightness and its standard deviation. We used the standard deviation instead of the (smaller) statistical uncertainty based on the number of detected events in order to capture the uncertainty induced by large-scale CXB variations, for example due to varying diffuse background emission in our Milky Way. The PIB level in the used 0.2(0.8)--2.3 keV energy band is always well below the CXB level. Two example surface brightness profiles for the sectors 0--180 deg and 180--360 deg, covering scales from $\sim$3 kpc to $\sim$1 Mpc, are shown in Fig.~\ref{fig:SB_prof1}. Given the large sectors, we used $30''$ bin widths to exploit the full spatial resolution offered by eROSITA.

\section{Results}
\label{sec:res}

\subsection{X-ray surface brightness distribution}
\label{sec:res:XSB}
\begin{figure}
    \centering
    \includegraphics[width=\hsize]{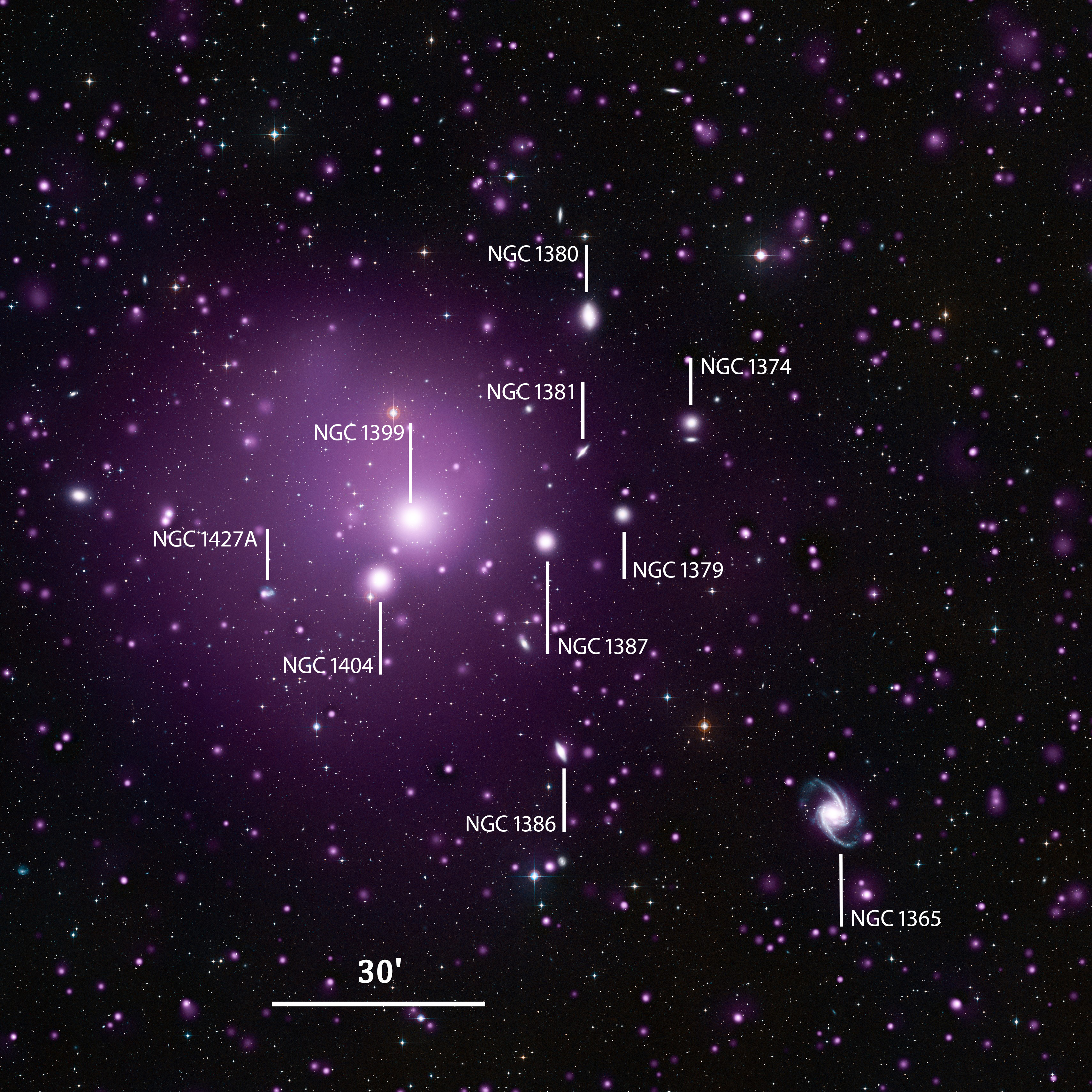}
    \caption{Digitized Sky Survey 2 image of the Fornax cluster in the optical B and R bands. The purple overlay is from the fully corrected eROSITA wavelet-filtered image, on an Asinh (hyperbolic arcsine) scale. The big elliptical galaxy close to the center of the image is NGC\,1399; the second brightest galaxy in the image about $13'$ south-southeast is NGC\,1404. Other prominent member galaxies are labeled in white. NGC\,1316, the central galaxy of the Fornax\,A group, is further to the southwest, outside the field shown here. Image credits: ESO and Digitized Sky Survey 2. Acknowledgment: Davide De Martin}
    \label{fig:opt_image}
\end{figure}

Focusing first on the core region of the Fornax cluster, we show in Fig.~\ref{fig:opt_image} the eROSITA surface brightness overlaid onto an optical color image obtained from the Digitized Sky Survey 2.\footnote{\url{https://www.eso.org/public/images/eso0949m/}}
The X-ray emission peak is centered on NGC\,1399. Several other of the bright Fornax member galaxies show X-ray emission, the brightest one being NGC\,1404. Furthermore, we observe characteristic features that could be due to sloshing as described in Sect.~\ref{sec:intro}. 

\begin{figure*}
    \centering
    \includegraphics[width=\hsize]
    {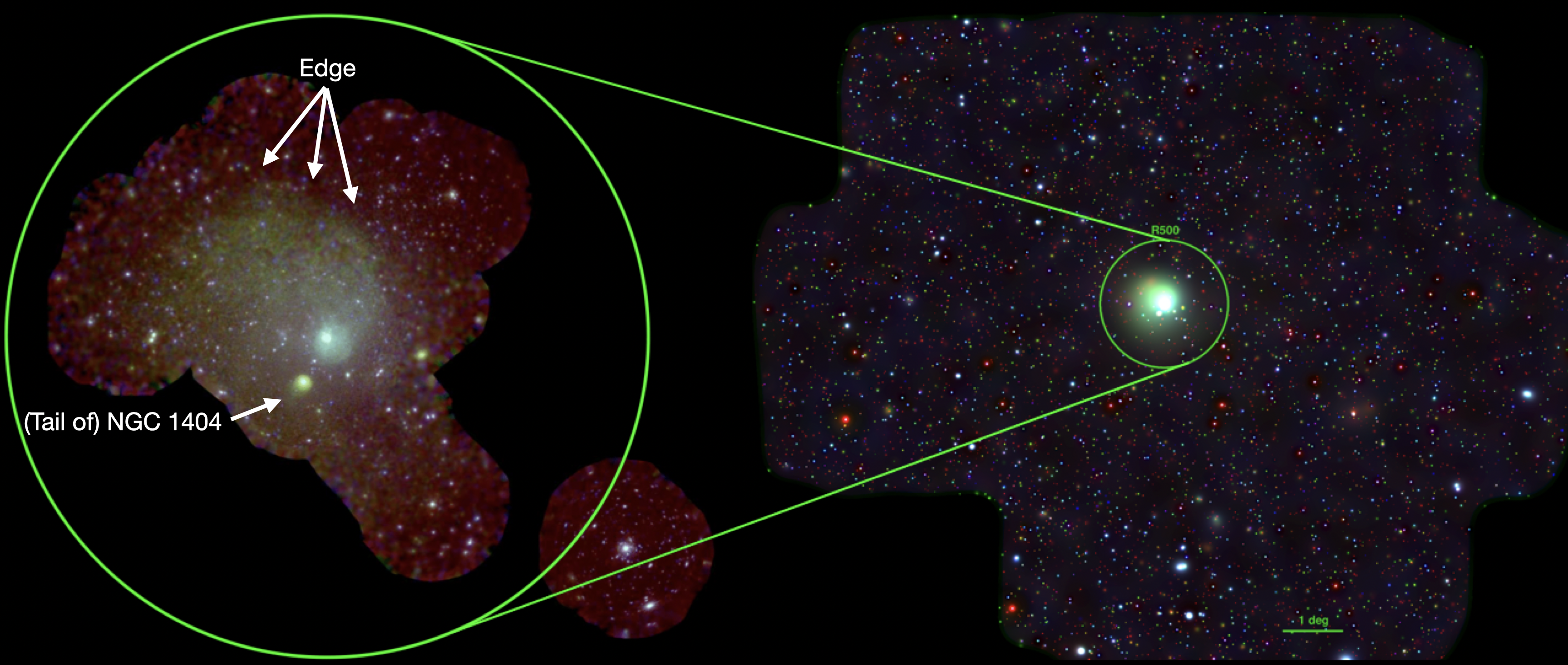}
    \caption{Right: Fully corrected, wavelet-filtered eROSITA RGB images in linear scale in the following energy bands: 0.2--1.0 keV (red), 1.0--1.4 keV (green), and 1.4--4.0 keV (blue). Left: \textit{XMM-Newton} mosaic of the central region (see Sect.~\ref{sec:res:XSB} for details). The large-scale ``edge'' spanning from the northeast to the northwest is labeled, as is NGC\,1404, whose X-ray tail points away from the Fornax cluster center.}
    \label{fig:RGB_lin}
\end{figure*}

To highlight these features more clearly, we show in Fig.~\ref{fig:RGB_lin} (right) a false color wavelet-filtered eROSITA RGB image based on individual fully corrected images in the energy bands 0.2(0.8)--1.0 keV, 1.0--1.4 keV, and 1.4--4.0 keV. The hot gas emission appearing in green sticks out in this representation compared to the large number of point sources (mostly active galactic nuclei in the background but also some foreground stars), which often appear redder or bluer. Furthermore, on the left we show a zoom into the core region using all the useful \textit{XMM-Newton} data available (1.6 Ms total flare-free exposure time of archival observations). The fully corrected and adaptively smoothed \textit{XMM-Newton} RGB image is constructed in the energy bands: red: 0.3--0.7 keV, green: 0.7--1.5 keV, blue: 1.5--4.0 keV. The same features and sources are visible in the two images, with \textit{XMM-Newton} showing more detail and sharpness in the innermost regions and eROSITA extending the field of view to beyond $R_{500}$.

\begin{figure}
    \centering
    \includegraphics[width=\hsize]{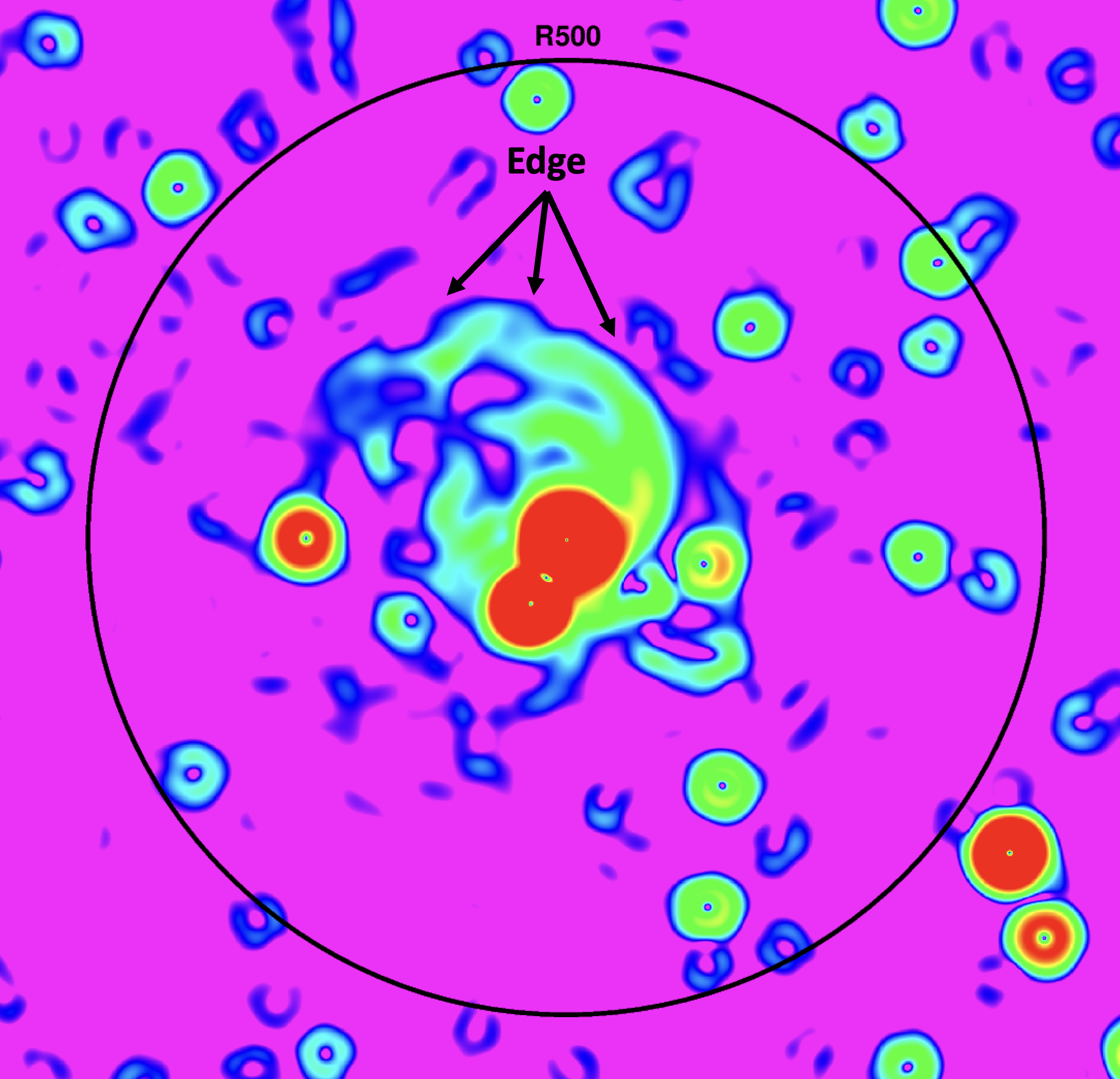}
    \caption{GGM-filtered eROSITA image in the energy band 0.2--2.3 keV, zoom into the central region, in log scale. The edge is prominent in this representation.}
    \label{fig:GGM}
\end{figure}

To further enhance the visualization of the position of the large-scale sloshing cold front (the ``edge''), we applied the Gaussian gradient magnitude (GGM) filtering technique \citep[][as implemented in \texttt{SciPy}, \citealt{2020NatMe..17..261V}]{2016MNRAS.460.1898S}, to the eROSITA image and show the result in Fig.~\ref{fig:GGM}. The similarity to the appearance of the edge in the 1.6 Ms \textit{XMM-Newton} image in Fig.~\ref{fig:RGB_lin} is striking.

Based on spectral analysis, \citet{2017ApJ...851...69S} identified the edge and the overall spiral pattern toward the inner regions as multiple sloshing cold fronts because the higher surface brightness regions at those contact discontinuities exhibit cooler temperatures. They infer the infall of NGC\,1404 as the likely cause of this pattern. The direction  of the NGC\,1404 tail discussed by \citep{2017ApJ...834...74S} and also seen in Fig.~\ref{fig:RGB_lin} fits this scenario, as do the hydrodynamical simulations described by \citet{2018ApJ...865..118S}.

The real power of eROSITA compared to all other X-ray telescopes in existence comes in through its sensitivity to large-scale low surface brightness features. Already the bottom image in Fig.~\ref{fig:image} revealed the vast extent of warm-hot gas emission, dwarfing the emission visible in the huge \textit{XMM-Newton} mosaic image shown in Fig.~\ref{fig:RGB_lin}. Note that we do \text{not} expect the surface brightness fluctuations in the bottom image in Fig.~\ref{fig:image} to be due to photon noise because of the wavelet-filtering process. Visual support for this expectation comes from comparing the eRASS:2 and eRASS3+eRASS4+eRASS5 images in Fig.~\ref{fig:erass345_12}: the two show essentially the same features despite coming from completely independent datasets. The quantitative analysis of the surface brightness profiles in different sectors follows below.

\begin{figure}
    \centering
    \includegraphics[width=\hsize]{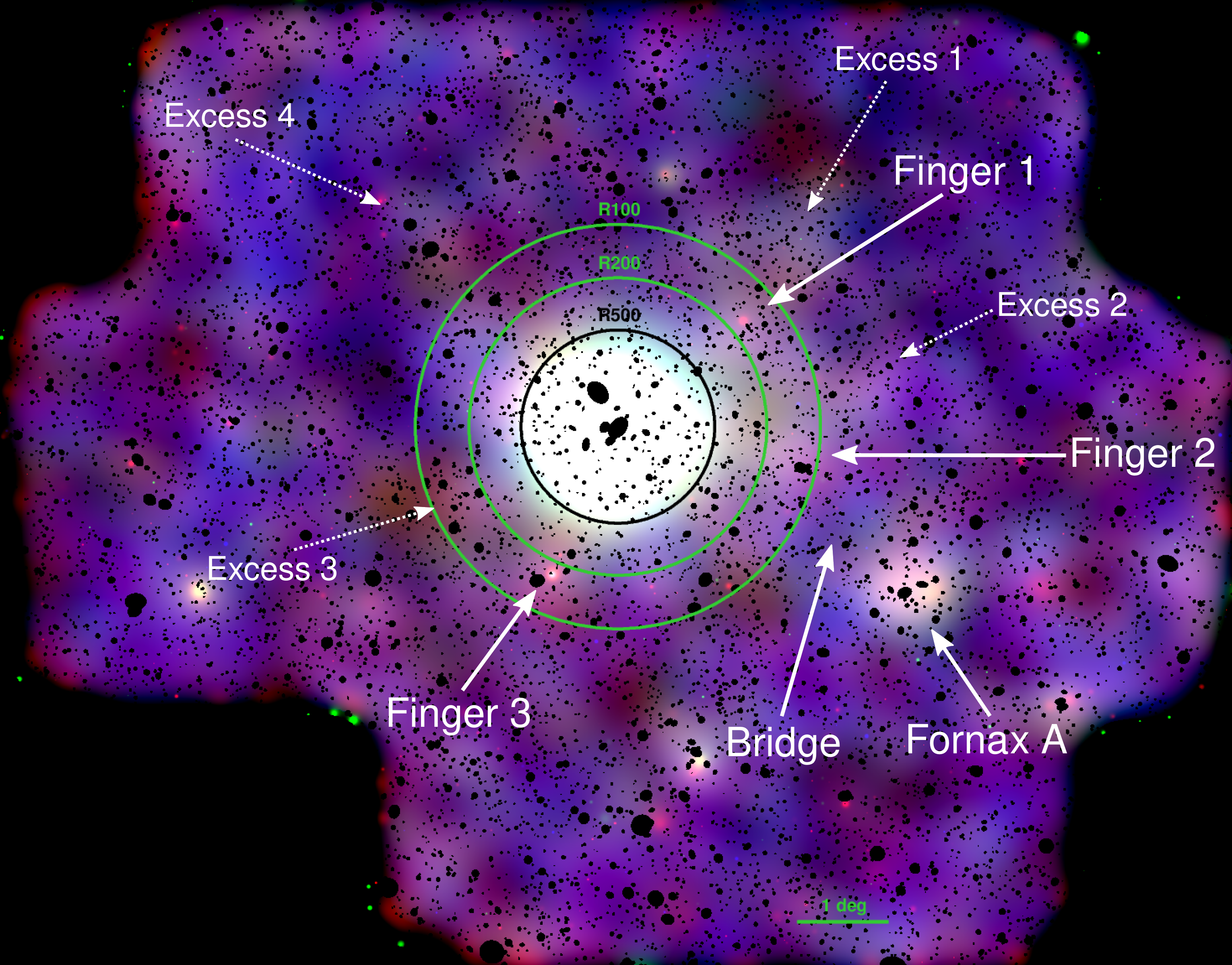}
    \caption{Same as Fig.~\ref{fig:RGB_lin} but on  a log scale and with sources of small extent excised. Scale limits were chosen to enhance low-surface-brightness fluctuations.}
    \label{fig:RGB_log}
\end{figure}

To make additional use of eROSITA's energy resolution, we show in Fig.~\ref{fig:RGB_log} an RGB image optimized to show low surface brightness fluctuations. Hot group and cluster gas emission is revealed in this image not only through its extent but additionally through its whitish color. Figures~\ref{fig:image} (bottom) and \ref{fig:RGB_log} both demonstrate that the Fornax cluster emission extends not only beyond $R_{500}$, already unreachable by the \textit{XMM-Newton} mosaic, but beyond $R_{200}$ and actually beyond even the virial radius, $R_{100}$, in some directions. In these images, we discover finger-like emission extensions beyond $R_{500}$ in particular toward the western and southeastern directions (labeled Fingers 1, 2, and 3 in Fig.~\ref{fig:image}, bottom). These extensions appear to stretch out to well beyond $R_{100}$, in particular Excesses 1 and 2.

On the other hand, we do not see any obvious evidence in the X-ray surface brightness maps of the shock front predicted by \citet{2018ApJ...865..118S} south of NGC\,1404. Below, we check quantitatively for features in the profiles in various sectors.
\begin{figure}
    \centering
    \includegraphics[width=0.8\hsize]{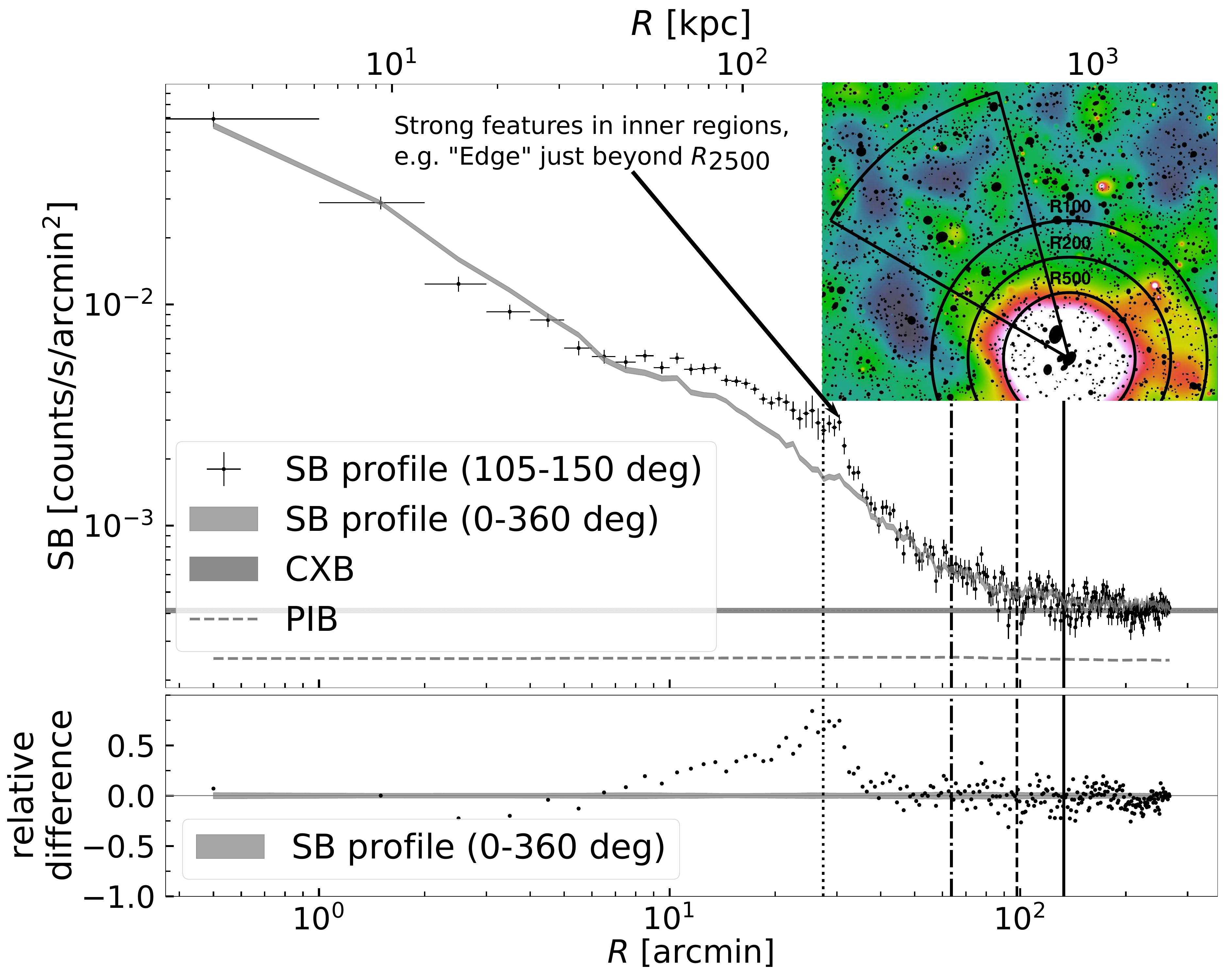}
    \includegraphics[width=0.8\hsize]{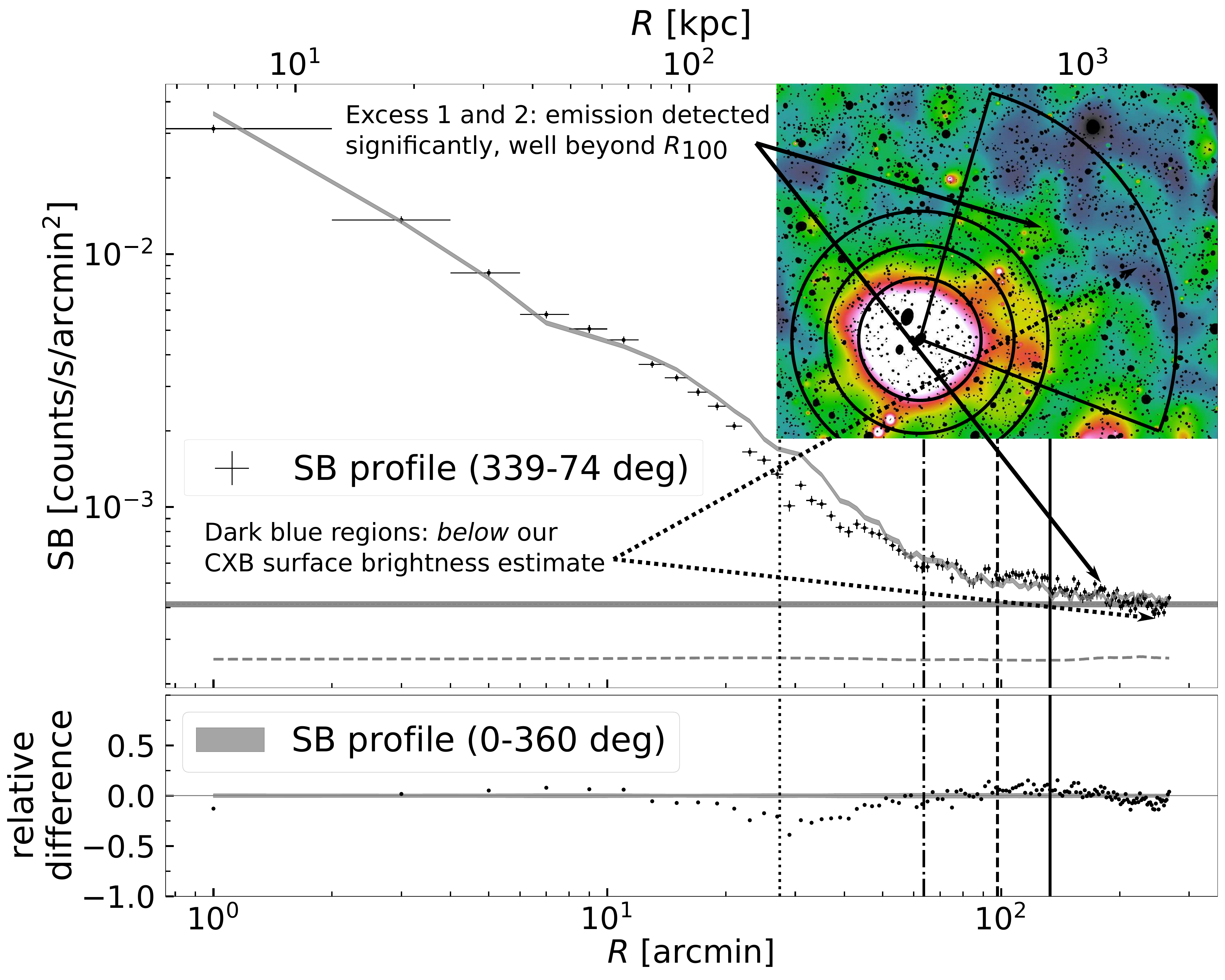}
    \includegraphics[width=0.8\hsize]{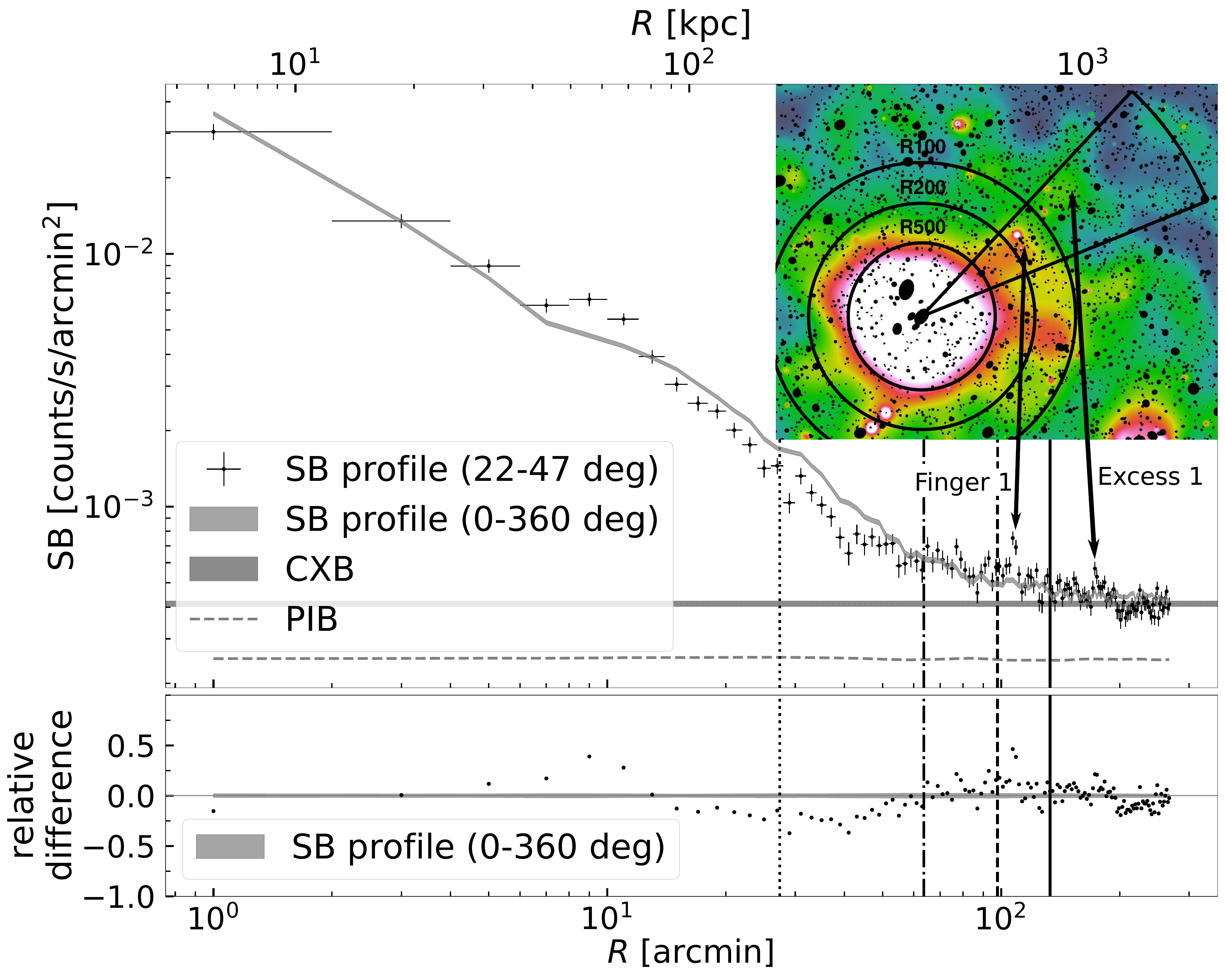}
    \includegraphics[width=0.8\hsize]{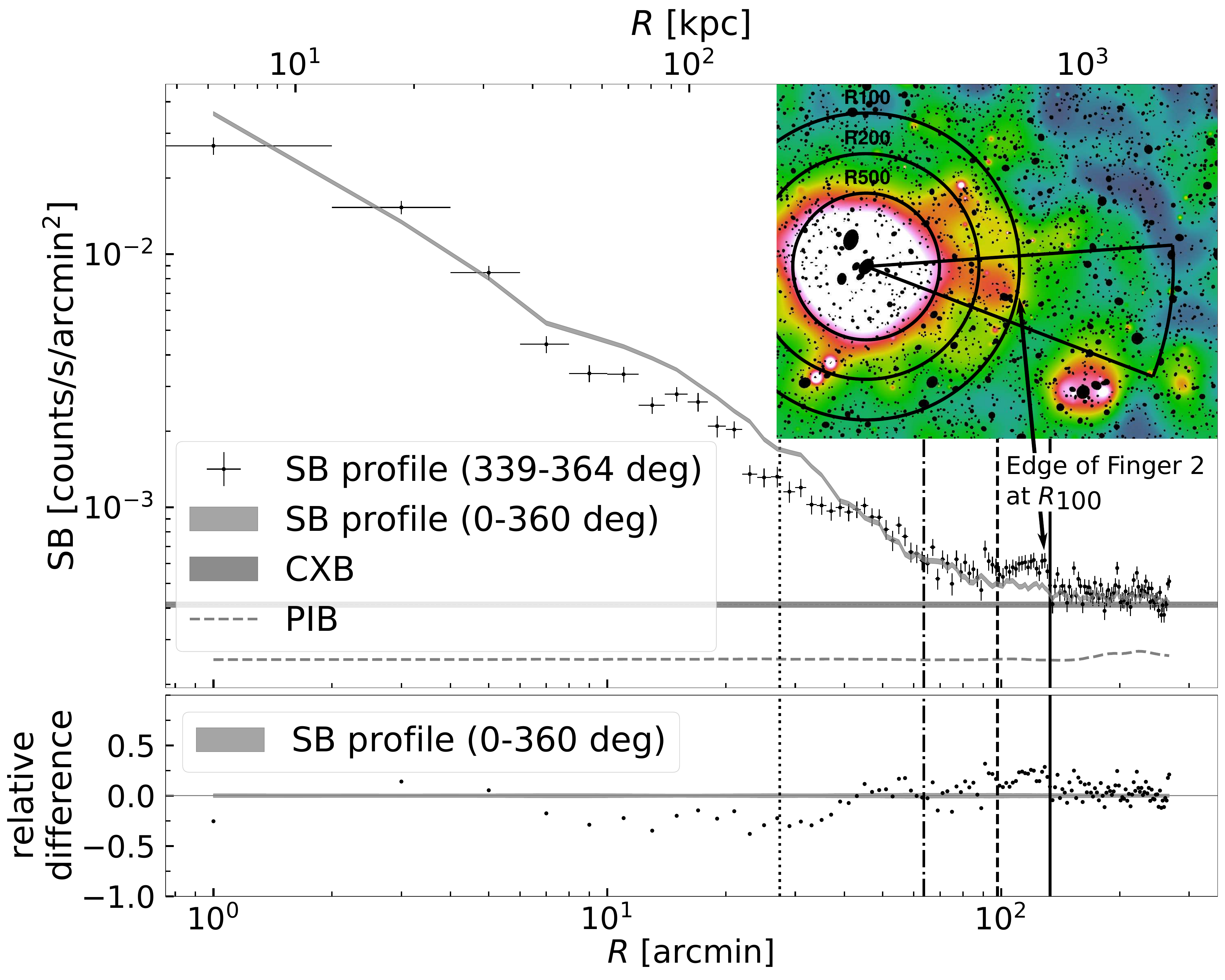}
    \caption{Surface brightness profiles for different sectors, with different features labeled.}
    \label{fig:SB_prof2}
\end{figure}

In principle, varying diffuse emission in our Milky Way (Fig.~\ref{fig:eRASS1_map}) or systematic uncertainties in the relative absorption correction can result in apparent surface brightness variations. Both effects would mostly be important at the softest energies $<$$1$ keV. Therefore, we constructed a hard band image (shown in Fig.~\ref{fig:hard_eRASS:5}) and checked if the emission features discussed above are robust. And, indeed, they are, lending confidence that they are not due to local effects.

The plots in Fig.~\ref{fig:SB_prof2} illustrate examples of surface brightness features discussed above.
A bin width of $60''$ was chosen for the uppermost plot as a compromise between the small sector and the relatively high surface brightness around $R_{2500}$; for the remaining three plots in Fig.~\ref{fig:SB_prof2} we chose a bin width of $120''$ as the focus lies on faint surface brightness features well beyond $R_{500}$. The 0--360 deg profile is always shown for comparison to highlight deviations from the azimuthal average profile; its bin width always matches that of the respective comparison profile.
The uppermost plot simply shows how well eROSITA can recover already known features in the inner regions; i.e., a strong drop of emission just beyond $R_{2500}$ -- part of the edge seen in Figs.~\ref{fig:RGB_lin} and \ref{fig:GGM}. The excess emission relative to the azimuthal average is highly significant ($\gg$$10\sigma$) as there are more than 10 bins above the average profile, each with a distance of several $\sigma$. The drop happens within about $240''$.

The next plot in Fig.~\ref{fig:SB_prof2} makes clear that Excesses 1 and 2 in the west-northwest beyond $R_{100}$ discussed above have, indeed, a significantly higher surface brightness than the CXB level as there are many data points consistently above this level. Quantitatively, there are $>$$10$ consecutive bins at least $1\sigma$ above the CXB level; i.e., the excesses are significant at the $>$$10\sigma$ level. Furthermore, it shows that our CXB level estimate is conservative because the surface brightness of the dark blue regions is even below our CXB level estimate.

The next two plots show how the surface brightness drops abruptly and significantly at the edges of Fingers 1 and 2.
Between $R_{500}$ and $R_{100}$ in the Finger 1 sector, almost all bins are well above the CXB level. Moreover, many of the bins also lie above the average profile. In particular, two consecutive bins between $R_{200}$ and $R_{100}$ lie several $\sigma$ above the average level; they mark the edge of Finger 1, potentially consisting of a clump of gas of width up to $\sim$$240''$ ($\sim$$25$ kpc).
Regarding Finger 2, between $R_{200}$ and $R_{100}$ there are $>$$10$ consecutive bins above even the average profile, almost all of them by a few $\sigma$. Therefore, the Finger 2 excess above average is detected at $>$$10\sigma$ significance.

\begin{figure}
    \centering
    \includegraphics[width=\hsize]{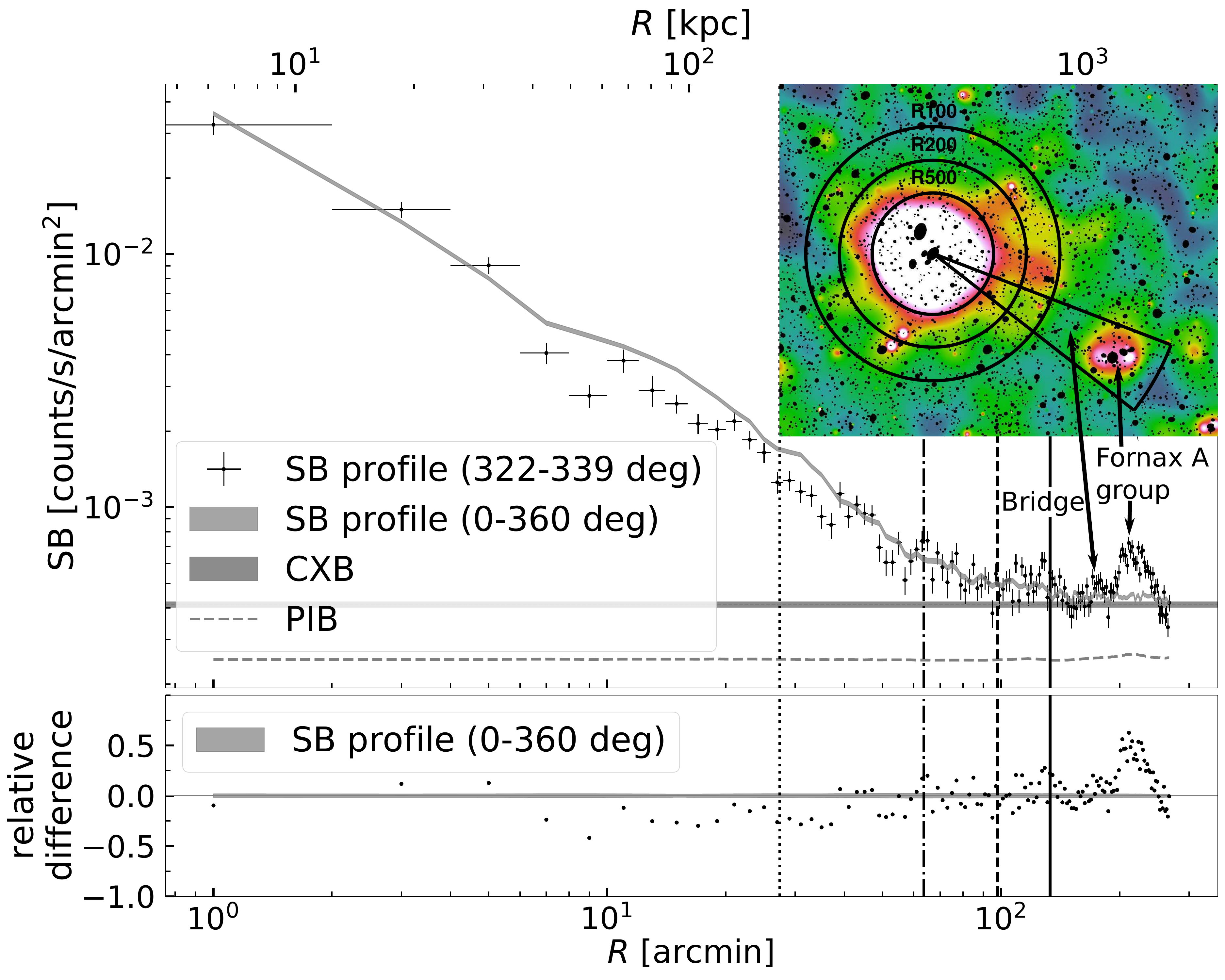}
    \caption{Surface brightness profile in the direction of the Fornax\,A group.}
    \label{fig:SB_prof3}
\end{figure}

We could then also search for the detached bow shock predicted by \citet{2018ApJ...865..118S} to lie 450--750 kpc south of NGC\,1399; i.e., roughly in the range between $R_{500}$ (393 kpc, 1.06 deg) and $R_{100}$ (816 kpc, 2.22 deg). Since it is not clear where exactly such a shock would be located and given the complexity of the two-dimensional emission in the outskirts, we started with a large sector and then looked at smaller ones in different directions. Figure~\ref{fig:SB_prof1} (bottom) shows the southern 180 deg sector. There are deviations from the average profile in the very inner region very roughly around $10'$ (here and also in many of the profiles shown in Figs.~\ref{fig:SB_prof2} and \ref{fig:SB_prof3}), which are related to the spiral sloshing pattern discussed earlier. However, no obvious surface brightness jumps that might be associated with a shock are obvious in the relevant radial range.
Figures~\ref{fig:SB_prof4} and \ref{fig:SB_prof5} show several more sectors. No obvious jump is present in the range 225--315 deg. In the 240--300 deg sector, there appear to be two peaks of emission surrounding $R_{200}$. However, consulting the surface brightness profiles of other sectors, in particular 225--270 deg and 240--270 deg, as well as Finger 3 toward the southeast in Fig.~\ref{fig:image}, it becomes clear that these are two extended (note that each individual bin covers 60$''$; i.e., twice the eROSITA point spread function) clumps of gas, not related to a shock front. On the other hand, the sector 180--270 deg does show a very weak indication of a surface brightness jump between $R_{500}$ and $R_{200}$; however, we consider it too weak, and the overall two-dimensional emission pattern too complex, to indicate a shock.

There are several possible reasons why no obvious bow shock is detected in the predicted range. On the simulation side, the set-up might be too idealized and, for example, without sufficient substructure, or it may not cover enough plausible merger scenarios. On the observational side, the signal might be insufficient for a significant detection, or the shape of the shock may not be well captured by our sector analyses.

The sector 270--360 deg in Fig.~\ref{fig:SB_prof5} shows a bump starting around $2R_{100}$, likely related to the Fornax\,A group. 
In Fig.~\ref{fig:SB_prof3} we then show the surface brightness in a small sector toward and beyond the Fornax\,A group. It is clear that significant group emission spans at least 1 deg. Moreover, the indication of an apparent emission bridge between the Fornax cluster and the Fornax\,A group is also confirmed in the surface brightness profile.

To get a better handle on the origin of these surface brightness extensions and excess regions, we compare them to other tracers of the large-scale structure in and around the Fornax cluster in the next sections.

\subsection{Distribution of cluster member galaxies}
\label{sec:res:gxs}
\begin{figure}
    \centering
    \includegraphics[width=\hsize]{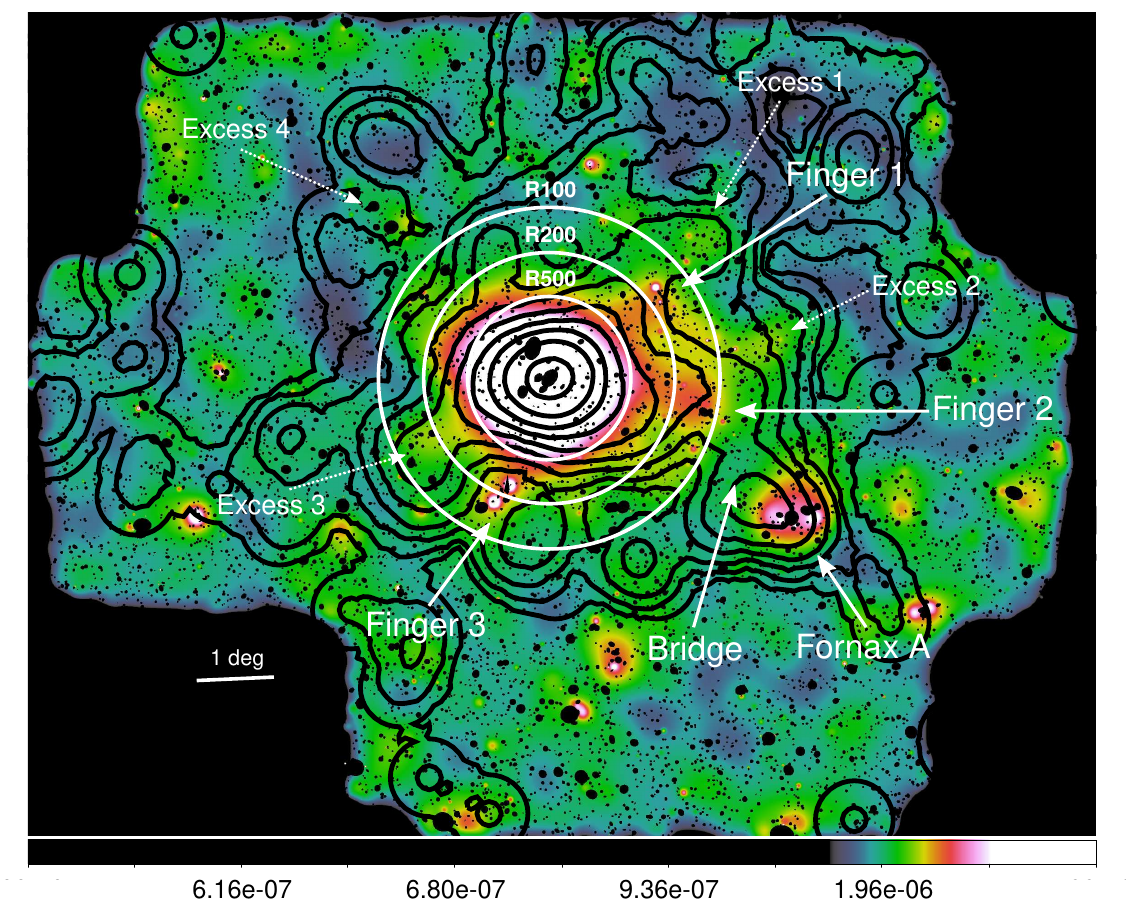}
    \caption{Same as Fig.~\ref{fig:image} but also showing density contours of galaxies at the Fornax cluster redshift from NED. Notice the overdensity of member galaxies in several of the regions of increased X-ray surface brightness in the outskirts, in particular to the west and southeast (i.e., the fingers and excesses) as well as the emission bridge to the Fornax\,A galaxy group.}
    \label{fig:image_gxs}
\end{figure}

The most obvious additional tracer of structure is the distribution of galaxies in and around the Fornax cluster.
As the Fornax cluster redshift is 0.005, we selected all objects classified as galaxies from the NASA/IPAC Extragalactic Database (NED)\footnote{\url{https://ned.ipac.caltech.edu}} in the image area with a redshift less than 0.01. All of these were assumed to be member galaxies. With this selection we aimed for a high completeness of member galaxies. Since velocity dispersion measurements for the Fornax cluster reach $\sigma_{\mathrm{los}}\approx 400$ km/s and above \citep[e.g.,][]{2001ApJ...548L.139D,2018MNRAS.481.1744P}, we wanted to ensure to include all galaxies within at least $3\sigma_{\mathrm{los}}\approx 1200$ km/s, which corresponds to a $z$ range of at least $0.001$--$0.009$. Figure~\ref{fig:image_gxs} shows the member galaxy distribution.

In the central region up to about $R_{500}$, the galaxy distribution exhibits an elongation in the east-west direction. On larger scales, well beyond $R_{500}$, one can notice a correlation of X-ray surface brightness excess regions and galaxy density. In particular, Fingers 1 and 2 and Excesses 1, 2, and 3 seem to have counterparts in the galaxy distribution. Moreover, the X-ray emission bridge toward the NGC\,1316 group also has an excess of galaxies;
this is related to the lopsided distribution of Fornax\,A group dwarf galaxies toward the northeast discussed in Sect.~\ref{sec:intro} and \cite{2021A&A...647A.100S}. The underlying reason for this offset between galaxy distribution and Fornax\,A intragroup gas might be related to the collisional gas lagging behind the collisionless galaxies in their common motion, for example due to ram pressure relative to the ICM of the Fornax cluster outskirts; however, a more detailed investigation of this scenario is beyond the scope of this paper.

\subsection{Distribution of intracluster ultra-compact dwarf galaxies}
\label{sec:res:dgs}
\begin{figure}
    \centering
    \includegraphics[width=\hsize]{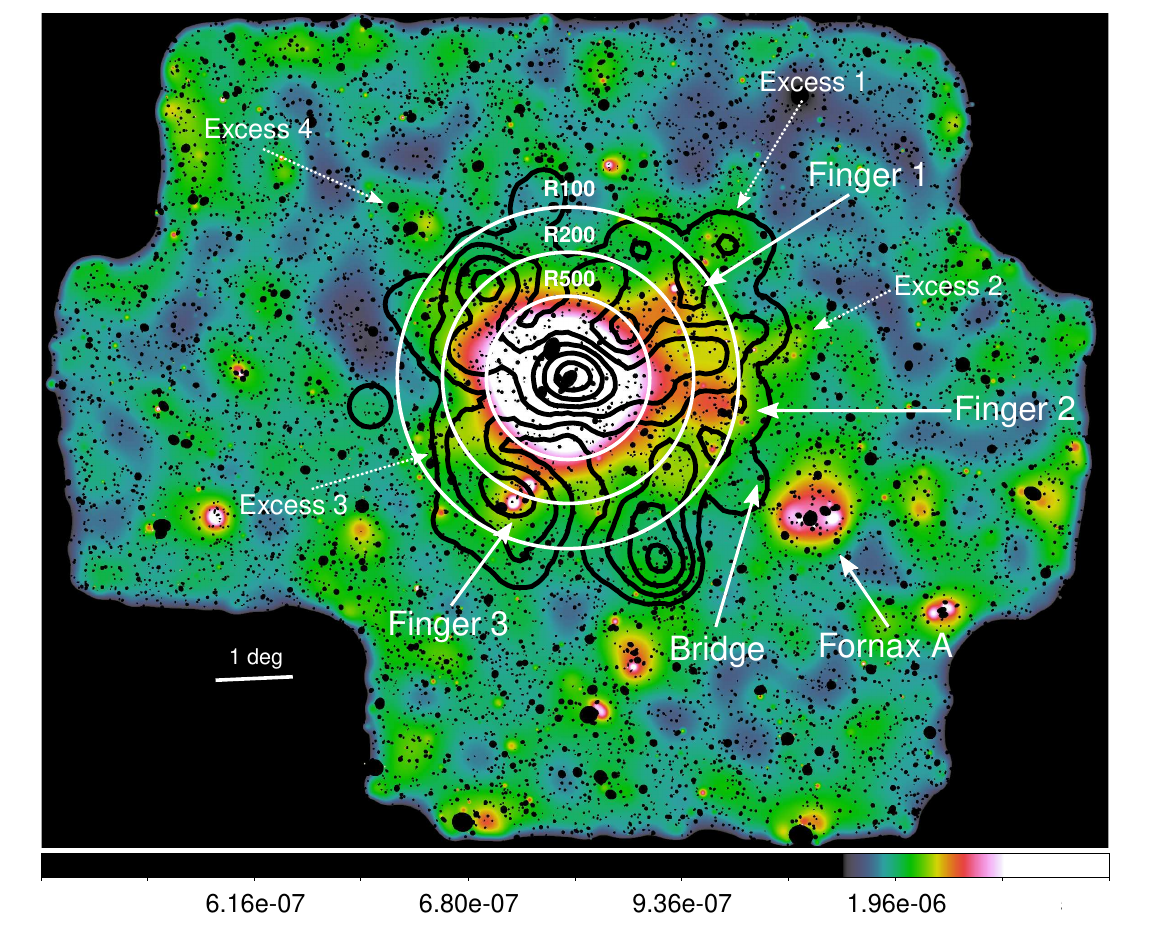}
    \caption{Same as Fig.~\ref{fig:image_gxs} but showing density contours of UCD galaxies in the Fornax cluster from \citet{2020A&A...639A.136C}. Notice the overdensity of UCD galaxies in regions of enhanced X-ray surface brightness in the outskirts, potentially marking preferred infall directions.}
    \label{fig:image_UCDs}
\end{figure}

As mentioned in Sect.~\ref{sec:intro}, \citet{2001ApJ...548L.139D} argued based on their higher velocity dispersion that dwarf galaxies might be preferentially infalling.
Tracers of disrupted dwarf galaxies are the so-called ultra-compact dwarf galaxies (UCDs), the most massive of them being stripped nuclei of formerly nucleated dwarf ellipticals \citep{1999A&AS..134...75H, 2003Natur.423..519D}.
In Fig.~\ref{fig:image_UCDs} we overlay the density of UCDs
from \citet{2020A&A...639A.136C} onto the eROSITA image. Overall, we find a broad correlation between the UCD densities and the X-ray surface brightness. In particular, Finger 3 sticks out while the correspondence to the other features is weaker. The elongation in the east-west direction in the central part also seems consistent with that observed in Fig.~\ref{fig:image_gxs}. 

\subsection{Distribution of intracluster globular clusters}
\label{sec:res:gcs}
\begin{figure}
    \centering
    \includegraphics[width=\hsize]{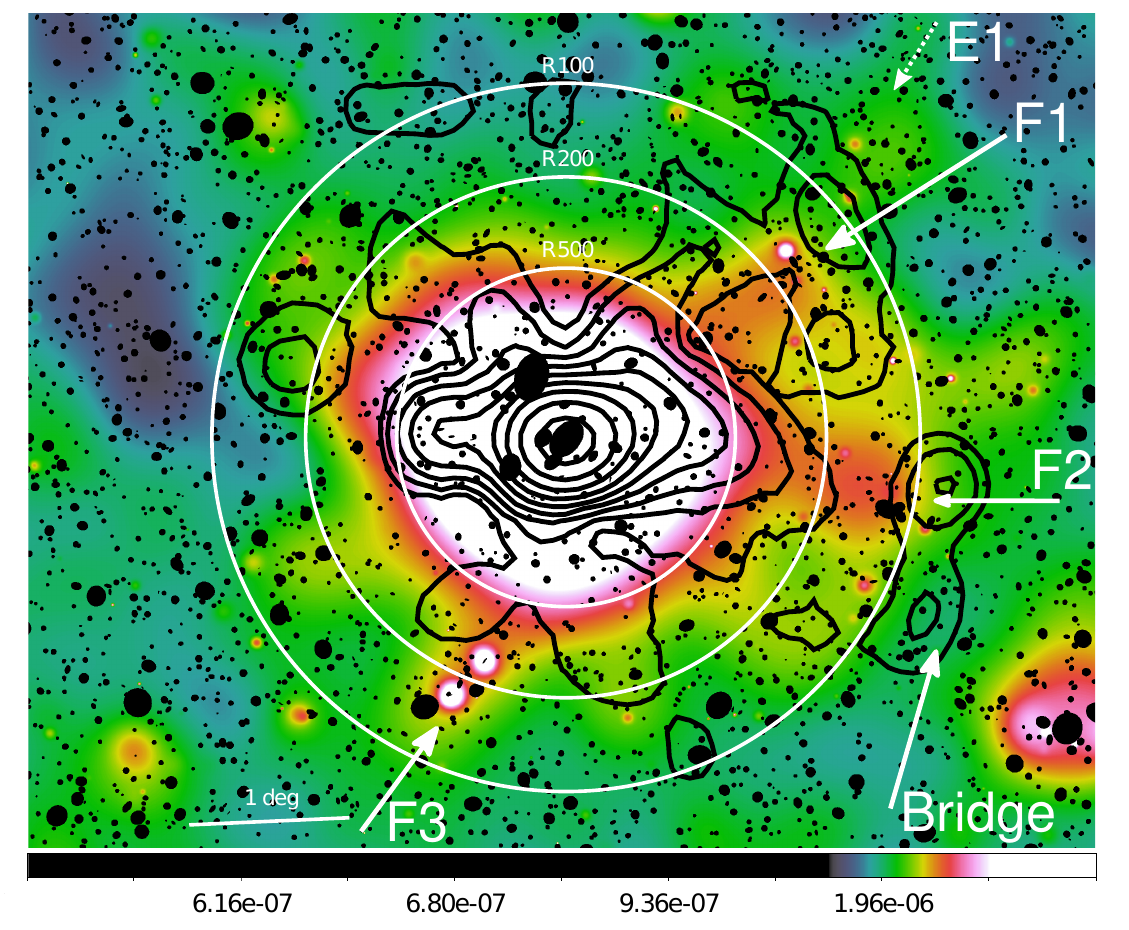}
    \caption{Same as Fig.~\ref{fig:image_gxs} but zoomed in more and showing density contours of globular clusters in the Fornax cluster from \citet{2020A&A...639A.136C}. The overdensities of globular clusters are in qualitative agreement with the  potential infall directions marked by UCDs and shown in Fig.~\ref{fig:image_UCDs}.}
    \label{fig:image_GCs}
\end{figure}

\citet{2020A&A...639A.136C} also presented the spatial distribution of
globular clusters in the Fornax cluster. The bright globular clusters in the inner region of $\sim$$1.5$ sq.deg around NGC\,1399 have been confirmed spectroscopically as well \citep{2022A&A...657A..93C}. The globular cluster density is shown in Fig.~\ref{fig:image_GCs} on top of the eROSITA image. Qualitatively, we find the same as for the UCD distribution -- east-west elongation in the center and a correspondence with the X-ray surface brightness in the outer parts. Especially Fingers 1 (F1) and 2 (F2), Excess 1 (E1), and the bridge seem to be traced approximately.

\subsection{Distribution of background clusters and superclusters}
\label{sec:res:cl}
\begin{figure}
    \centering
    \includegraphics[width=\hsize]{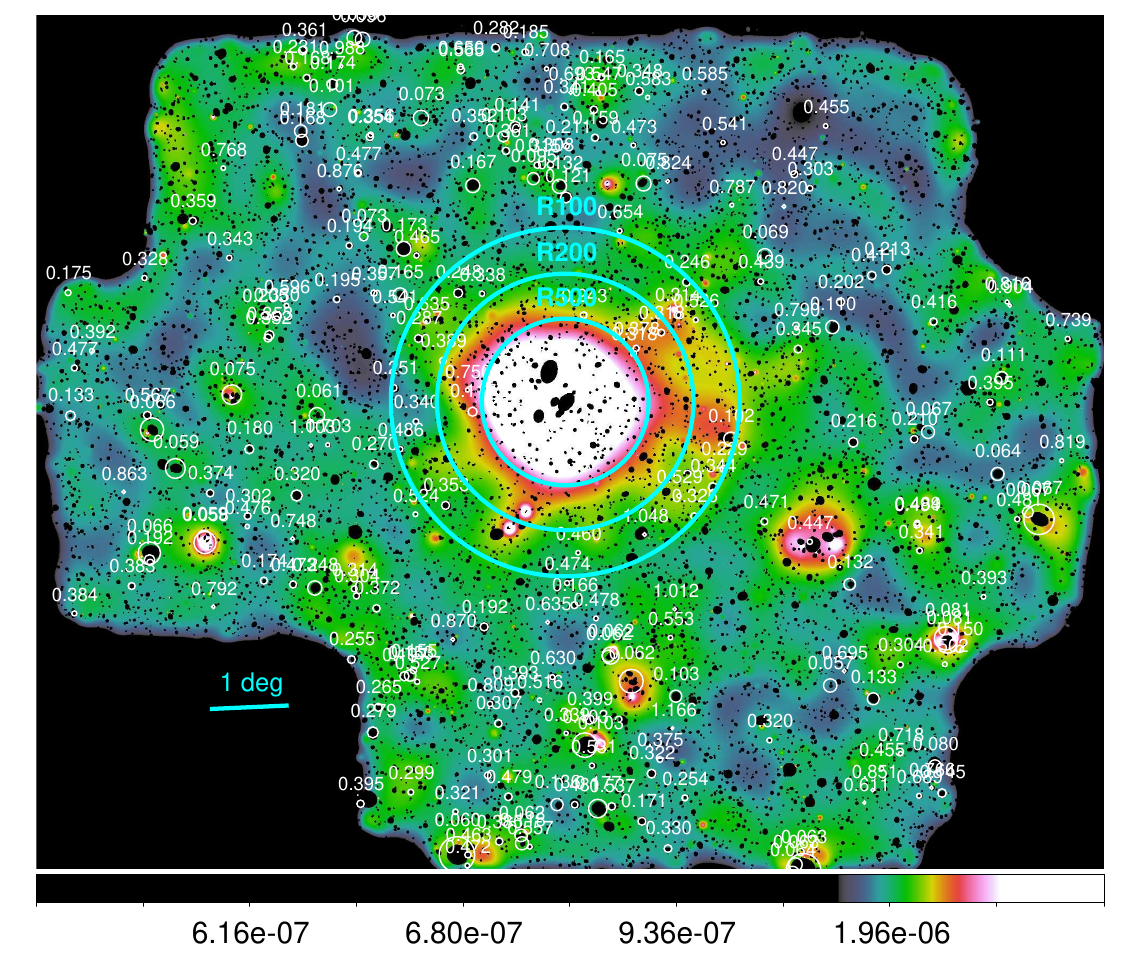}
    \caption{Same as Fig.~\ref{fig:image} but also showing the positions of background clusters from the eRASS1 catalog \citep{2024arXiv240208452B,2024arXiv240208453K}. Redshifts are labeled, and the circle radii correspond to $R_{500}$. See Figs.~\ref{fig:erass1_zoom1} and \ref{fig:erass1_zoom2} for zoomed-in versions.}
    \label{fig:erass1}
\end{figure}

eROSITA has discovered more than 12,000 galaxy clusters in more than 13,000 sq.deg across the western Galactic hemisphere \citep{2024arXiv240208452B,2024arXiv240208453K}, including the region of the Fornax cluster. Therefore, we can expect about one background cluster per square degree, resulting in a large number to lie in the field under study here. As described in Sect.~\ref{sec:red}, we excised compact sources, which we expected to include background clusters as most of them will be at comparatively high redshifts, $z\gtrsim 0.1$ (the median redshift of the cluster catalog is 0.31), resulting in a small apparent size. Still, if there was an overdensity of clusters in a certain area, either just projected along the line of sight or forming a real supercluster, there might be significant summed emission present from their un-excised outskirts or from filaments connecting them. We tested this in order to avoid incorrectly assigning excess emission to the outskirts of the Fornax cluster. To do so, we overlay in Fig.~\ref{fig:erass1} the eRASS1 background clusters. Their redshifts are labeled and the circle sizes correspond to their $R_{500}$ radii.

Several things can be noted in this figure: (i) Our source detection indeed excises almost all of the background clusters, with a region size that is often comparable to $R_{500}$. While the source detection process here is very different, this is nonetheless not surprising given that we used data with five times the exposure time compared to eRASS1. It is also not surprising that there is a small fraction of clusters that we do not detect because (a) the purity of the eRASS1 catalog is estimated to be 86\% \citep{2024arXiv240208452B}
and (b) a few of them are very nearby ($z\lesssim 0.05$) and, therefore, not compact. (ii) Some of the regions of enhanced surface brightness in the field indeed do correspond to overdensities of clusters, sometimes from a few at the same redshift, sometimes at different redshifts. While the clusters themselves are essentially all excised, there appears to be emission well beyond their $R_{500}$ radii that, combined, is picked up by eROSITA. The cause of this apparent large-scale structure emission (e.g., WHIM surrounding clusters or an overdensity of undetected galaxies, active galactic nuclei, or small groups of galaxies surrounding the clusters) will be studied in a future paper. (iii) Most of the main excess regions described in previous sections do not correspond to such cluster overdensities, making it unlikely they are due to projection of emission from the outskirts of background clusters. (iv) There are, however, a few X-ray-excess regions that do show background cluster overdensities. Particularly interesting is that there are three clusters at the same redshift ($\sim$0.31) aligned almost perfectly with Finger 1 that points in the northwest direction (projected roughly at the Fornax $R_{200}$). The projected extent is about 15 Mpc, in principle reasonable for a filament \citep{gcgws24}. Nonetheless, it seems unlikely that this emission finger is due to a background filament, especially because it would have to be a rather bright and thick ($\sim$10 Mpc) filament. Indeed, its surface brightness is comparable to that of the few rare filaments individually detected in X-rays \citep{dpr24,rvp21}, which are all at least five times less distant, and the typical width of a filament in numerical simulation is of order 1\,Mpc \citep{galt21,wwgkl24}. Therefore, we conclude that this particular alignment of three projected background clusters with Finger 1 is a coincidence and that their outskirts or filament emission is not responsible for a large fraction of the observed X-ray excess.

In summary, we do find some hints for diffuse emission associated with the outskirts of background clusters but conclude that none of the features discussed here (fingers and excesses) are caused by them.

\subsection{HI-tail galaxies}
\label{sec:res:HI}

Cold gas stripping is widespread for galaxies falling into galaxy clusters \citep[e.g.,][]{2021PASA...38...35C}. Whether the same processes happen in lower-mass galaxy groups is less clear. With a blind ATCA HI survey, \citet{2021A&A...648A..31L} did find evidence of neutral hydrogen removal within the virial region of the Fornax cluster. Using MeerKAT, \citet{2023A&A...673A.146S} recently increased the number of known HI-tail galaxies in Fornax from one to six, showing that ram pressure stripping can operate in this group, especially for galaxies that underwent interactions, thereby making their gas less bound. Four out of these six are located at projected radii beyond $R_{500}$, thereby making their X-ray environment inaccessible to \textit{XMM-Newton} or \textit{Chandra}.
\begin{figure}
    \centering
    \includegraphics[width=\hsize]{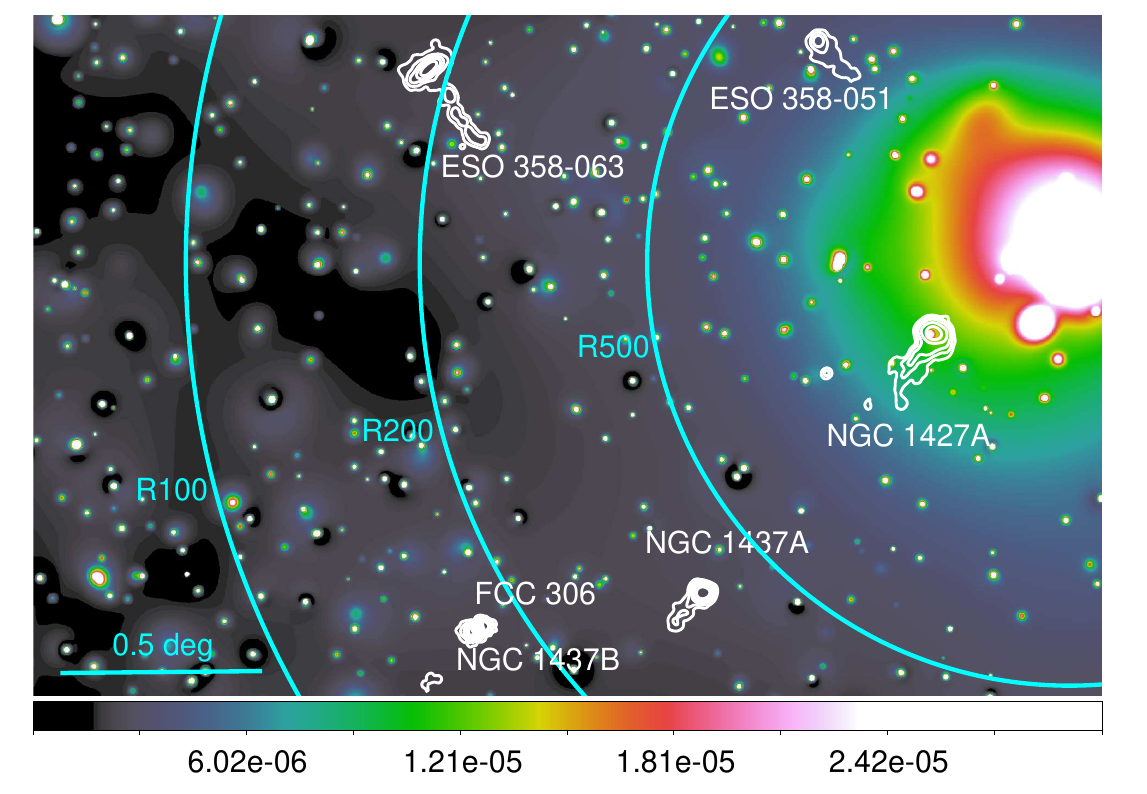}
    \caption{Zoom into the eROSITA image shown in Fig.~\ref{fig:image}, top, plus atomic neutral hydrogen emission contours from MeerKAT from \citet{2023A&A...673A.146S}, illustrating ram pressure stripping by the Fornax cluster ICM in action.
    }
    \label{fig:MeerKAT_HI}
\end{figure}

In Fig.~\ref{fig:MeerKAT_HI} we show the HI contours from \citet{2023A&A...673A.146S} on top of the eROSITA image. None of the host galaxies themselves shows strong X-ray emission. All four galaxies are in a region where X-ray emission from the Fornax ICM is clearly detected by eROSITA, though. Intriguingly, the HI tails for three of them (FCC\,306, NGC\,1437B, and NGC\,1437A) are parallel to the southeastern Finger 3 and coincident with an overdensity of possibly infalling UCDs (Fig.~\ref{fig:image_UCDs}). The tail of NGC\,1427A located inside $R_{500}$ also points in the same direction. All of their tails point away from the center of the Fornax cluster, consistent with them falling in for the first time, loosing their HI gas along the way due to ram pressure stripping despite the relatively low ICM density in the Fornax outskirts. Furthermore, the tails of the two northern galaxies, ESO\,359-063 and ESO\,358-051 align with the direction toward Excess 4 and also seem coincident with a UCD overdensity, although their tails point away from the Fornax center. We conclude that, in the Fornax cluster, HI gas is being stripped from galaxies that are (still) far away from the cluster center and that are moving along preferred infall directions; this might also have implications for ICM heavy element enrichment processes.

In this context, it is interesting to note that 
\citet{2024ApJ...969...28C}
have shown, by analyzing Fornax cluster-like halos in the Illustris TNG50 simulation, that stripped HI gas should exist out to beyond the virial radius, with a non-negligible covering fraction.

\subsection{NGC\,1316}
\label{sec:res:ngc1316}
\begin{figure}
    \centering
    \includegraphics[width=\hsize]{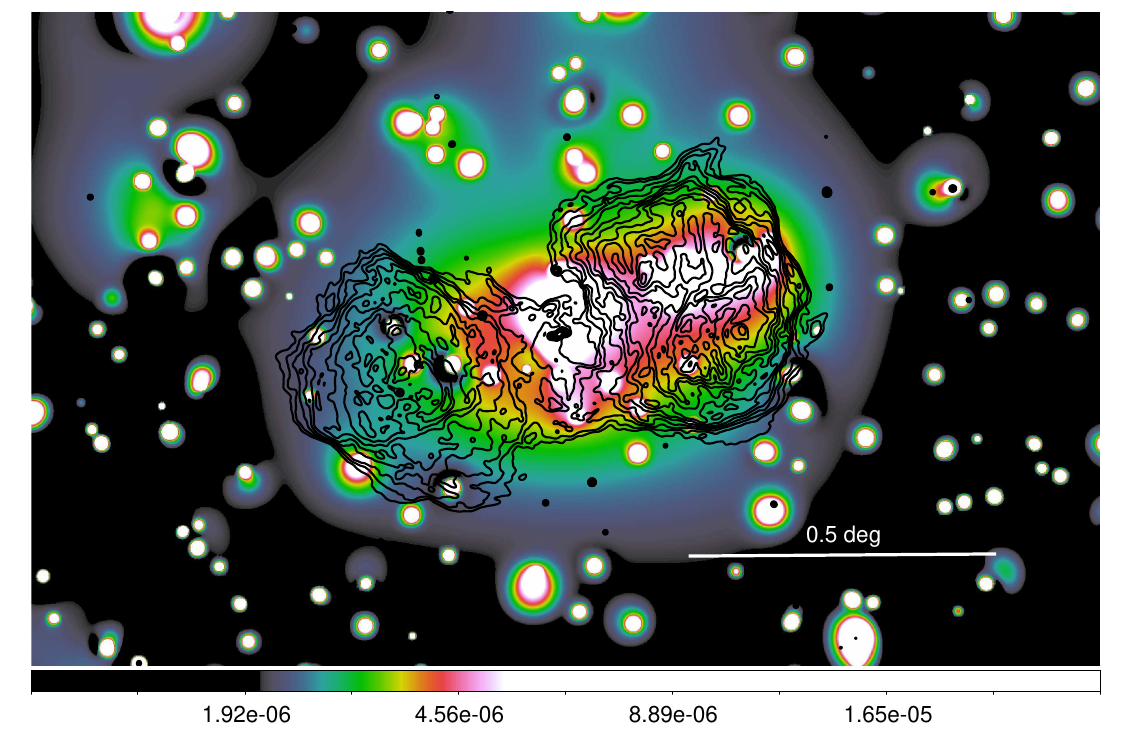}
    \caption{Fully corrected, wavelet-filtered image in the energy band 0.2--2.3 keV zoomed in on the Fornax\,A radio galaxy (NGC\,1316). Overlaid are MeerKAT 1.44 GHz contours from \citet{2020A&A...634A...9M}.
    Note that the diffuse X-ray emission is aligned with the radio jet and lobe direction but extends farther, for a total linear size of at least 1 deg.}
    \label{fig:MeerKAT}
\end{figure}

NGC\,1316 is the central galaxy of a galaxy group that appears to be bound to the Fornax cluster and falling into it \citep[][]{2001ApJ...548L.139D}. It also hosts Fornax\,A, a bright radio active galactic nucleus exhibiting very extended lobes. In Fig.~\ref{fig:MeerKAT} we show the eROSITA X-ray surface brightness and overlaid on it the MeerKAT radio contours from \citet{2020A&A...634A...9M}. One can notice that the elongation of the X-ray surface brightness aligns with that of the radio lobes. Furthermore, we see in this image that the X-ray emission extends much farther than the radio lobes. The total extent is clearly more than 1 deg. Actually, looking at Fig.~\ref{fig:image}, bottom, the total extent might well be 2 deg in this direction. This large extent, as well as the fact that the X-ray emission does not trace the radio lobes exactly, shows that the soft X-ray emission traced by eROSITA
is dominated by thermal emission from the NGC\,1316 intragroup gas instead of by inverse Compton scattering of the photon background on the relativistic electron population that gives rise to the synchrotron emission observed in the radio regime.
This appears to be in contradiction to earlier studies that assume that at 1 keV all X-ray emission has a nonthermal origin \citep[e.g.,][]{2019MNRAS.485.2001P}. At higher energies $\gtrsim$1.5 keV, the nonthermal component has been shown by \textit{Suzaku}   spectral analysis to dominate instead \citep[e.g.,][]{2009PASJ...61S.327T}.

Given the emission bridge that connects NGC\,1316 with the Fornax cluster that we have discovered (see Fig.~\ref{fig:image_gxs}), it actually  becomes hard to determine at all where the NGC\,1316 group emission ends and where the Fornax cluster emission begins.

\subsection{SLOW comparison}
\label{sec:res:SLOW}
The simulation of the local Universe (SLOW) simulations \citep{2018MNRAS.478.5199S, 2023A&A...677A.169D} aim to reproduce the local large-scale structure by convolving randomly generated initial conditions with observed large-scale structure data \citep{2016MNRAS.455.2078S}. A direct consequence is a set of simulated counterparts for locally observed galaxy clusters \citep{2024arXiv240201834H} reproducing positions and general properties with good accuracy. Recent comparisons to observations \citep[e.g.,][]{dpr24} have especially shown that the large-scale environment components of matched clusters such as massive neighbors and filaments match closely due to the reliable constraints on these scales. To test whether this holds for the relatively low-mass Fornax cluster, we produced a mock observation of the simulated counterpart of the Fornax cluster in the SLOW simulation. We used the post-processing program SMAC \citep{2005MNRAS.363...29D} to simulate X-ray emission in the 0.1--2.4 keV band.
Additionally, we optimized the observer position for the mock observation so that Fornax was in exactly the correct sky position in order to eliminate projection uncertainties generated by the slight offsets from the true positions the simulation naturally generates.
\begin{figure}
    \centering
    \includegraphics[width=\hsize]{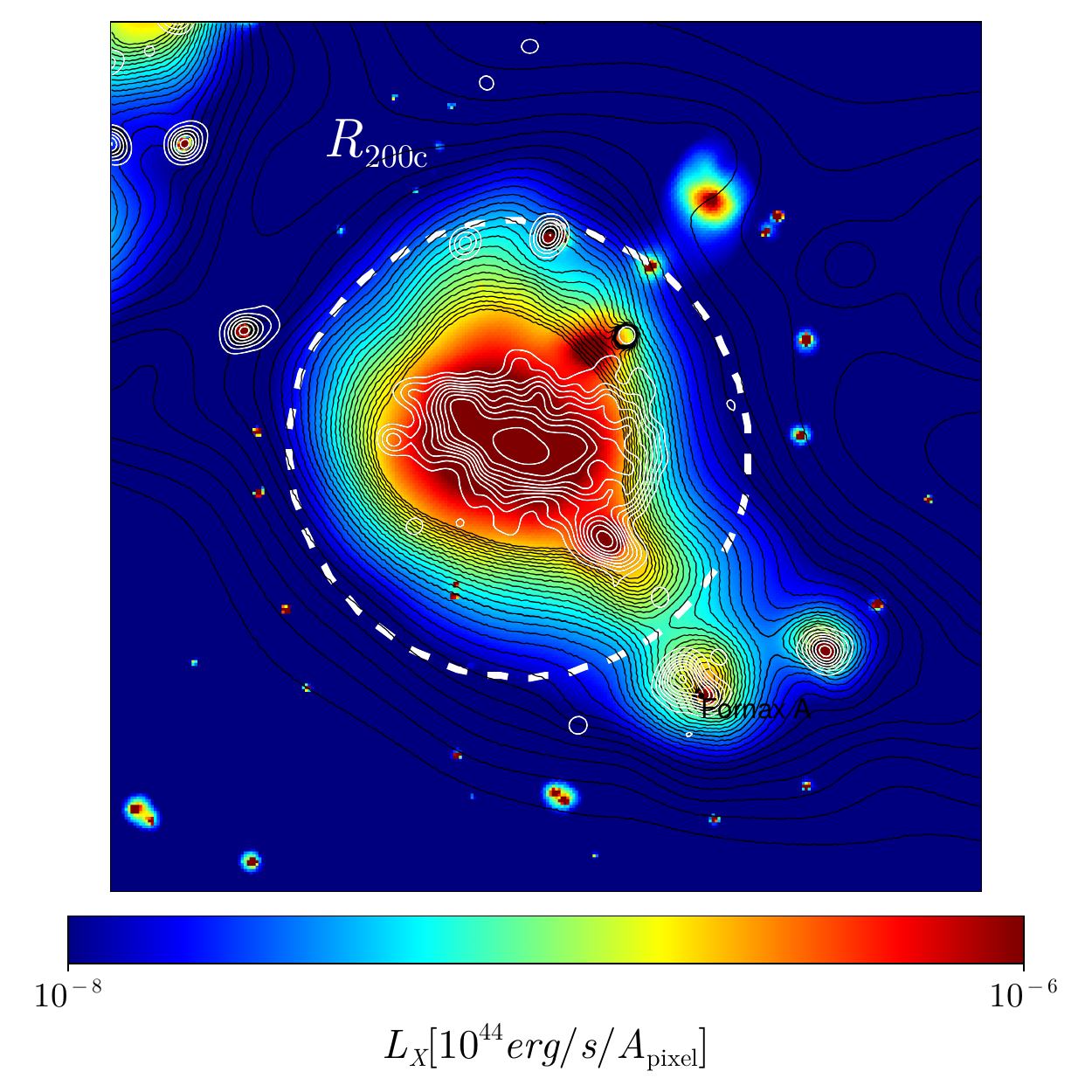}
    \caption{ Fornax cluster and the bridge to Fornax\,A as they appear in the X-ray band (0.1--2.4 keV) SLOW mock (projection out to 350 Mpc). Fingers 1 and 2, as seen in the eROSITA image, are also present (but see the main text). The black contours indicate the gas density levels in the interval $\rho=[0.8,6.0]\times \rho(R_\mathrm{200})$ in a 25 Mpc slice centered on the Fornax cluster. Similarly, the white contours show stellar density in the same slice. Based on \citet{2023A&A...677A.169D} and \citet{2024arXiv240201834H}.}
    \label{fig:SLOW}
\end{figure}

In Fig.~\ref{fig:SLOW} we show the result of the SLOW simulation \citep{2023A&A...677A.169D,2024arXiv240201834H} at the position of the Fornax cluster. Comparison to Fig.~\ref{fig:image} shows, interestingly, that the bridge toward Fornax\,A is also present. Furthermore, emission fingers to the west can be clearly seen in the image, which resemble Fingers 1 and 2 in the eROSITA image. Scrutinizing the simulation data, it turns out that, indeed, the finger to the southwest (Finger 2) is colocated with the Fornax cluster. However, the emission in the image corresponding to Finger 1 is actually located 53 Mpc behind the Fornax cluster and is due to a projected galaxy group. When looking in detail at the black gas density contours, though, a weak higher-density extension becomes obvious toward the northwest, right where the eROSITA image shows Excess 1. The white contours show the stellar density, and the similarity to the observed galaxy and globular cluster density (Figs.~\ref{fig:image_gxs} and \ref{fig:image_GCs}, respectively) is striking. First, the same elongation in east-west direction is seen. Second, the offset of the galaxy and gas distributions close to the Fornax\,A group (discussed in Sect.~\ref{sec:res:gxs}) is also present. Analyzing the galaxy velocities in the SLOW simulation shows that those galaxies, as well as those close to Finger 2, are currently moving toward the Fornax cluster center. In conclusion, the agreement between the SLOW simulations and the eROSITA and galaxy observations, on the one hand, lends confidence to the realism of the constrained simulations and, on the other hand, robustly aids the interpretation of the observations. 

\section{Discussion and conclusions}
\label{sec:dis}
With the new eROSITA observations, we have discovered X-ray emission from the Fornax cluster beyond $R_{100}$ for the first time. In the inner parts, $<$$R_{500}$, we detect the large-scale sloshing features already known from previous \textit{XMM-Newton} mosaic observations. In the outer parts, the emission is not symmetric but instead shows preferred directions of higher X-ray surface brightness, as one might expect from gas infall along filaments. In particular, we find finger-like structures with sharp surface brightness drops around $R_{100}$, which then slowly fade out to much larger distances as regions of excess emission.

At different levels, the NED member galaxy, UCD galaxy, and globular cluster distributions are all correlated with the X-ray-excess regions, and the HI tails of some member galaxies also point in the same preferred directions. This raises the intriguing possibility that we are witnessing the ongoing growth of the Fornax cluster along preferred infall directions, a growth that matches that of the Fornax counterpart in the SLOW simulation.
The cluster assembly is still ongoing along filaments where preprocessed galaxy groups (with or already without their own X-ray gas) reach the cluster core in specific directions. In the Fornax cluster, this led to an east-west elongation of the galaxy, UCD, and globular cluster distributions (Figs.~\ref{fig:image_gxs} through \ref{fig:image_GCs}). In these regions UCDs trace  dissolving, nucleated galaxies, with some globular clusters still related in phase space to the accretion event (e.g., \citealt{2022A&A...657A..94N} and \citealt{2022A&A...657A..93C}).
Since intracluster globular clusters are also related to the intracluster light and dark matter distribution in galaxy clusters, all these tracers map the mass assembly of the galaxy cluster, as does the X-ray gas \citep[e.g.,][]{2012MNRAS.425...66F,2023A&A...679A.159D,2024ApJ...975...76M}. All these tracers in the Fornax cluster therefore seem to outline the preferred infall directions.
Alternatively, at least some of these features might be related to the ongoing minor merger with NGC\,1404. However, we do not find obvious evidence of the shock front predicted by numerical simulations of this merger \citep{2018ApJ...865..118S}.

It is worth noting that no perfect agreement among the large-scale structure tracers may necessarily be expected. For example, for the NED galaxy distribution no distinction is made here between newly infalling galaxies and galaxies that are already in full virial equilibrium with the gravitational potential.
The UCD galaxy distribution, on the other hand, might be dominated by infalling galaxies. HI tails clearly demonstrate infall, and the projected direction is clear as well.
Last but not least, for the X-ray surface brightness, all corrections have been performed. However, when comparing the surface brightness distribution with the distributions of the other tracers, one needs to be cautious because for none of the tracers based on individual objects (NED galaxies, UCDs, globular clusters, or HI-tail galaxies) have any selection effects been taken into account. For example, as the NED archive is not based on a complete catalog, no selection effect correction of galaxy density estimates can be performed. Therefore, there is uncertainty as different areas might have different sensitivities to the various tracers.

We have furthermore discovered a bridge of X-ray emission toward NGC\,1316, which hosts Fornax\,A. This bridge approximately coincides with a region of enhanced Faraday depth as measured by \citet{2021PASA...38...20A}, which itself also indicates the presence of a significant number of thermal electrons (and magnetic fields). The X-ray emission of the galaxy group around NGC\,1316 itself extends well beyond 1 deg and is aligned with the radio lobes, indicating interaction between the relativistic and thermal electrons.

The detection of higher X-ray surface brightness in the outskirts of background clusters in overdense regions of the field implies that with eROSITA we start going beyond the detection of clusters as individual objects and instead see a Universe of continuous large-scale emission -- as expected from cosmological simulations.

Based on the hard band image (1.0--2.3 keV), we are confident that most of the X-ray features are not due to soft X-ray structure in our own Milky Way; still, it remains difficult to confidently rule out the possibility that some of the X-ray-excess regions are due to foreground structure (see, e.g., the extreme cases of emission of the eROSITA bubbles projected onto the outskirts of the Virgo and Centaurus clusters; \citealt{mrv24,vrp24}). Correlations between X-ray-excess regions and various other tracers of structure, such as member galaxies and intra\-cluster globular clusters, as employed here do support our conclusions. Final confirmation of the distance of individual excess regions may need to await high-spectral-resolution follow-up observations with XRISM \citep[e.g.,][]{2020arXiv200304962X}, HUBS \citep[e.g.,][]{2023SCPMA..6699513B}, or even NewAthena \citep[e.g.,][]{nbb13,2025arXiv250103100C}, from which one will be able to determine the redshifts of the excess regions from their X-ray emission lines.

\begin{acknowledgements}
We thank the anonymous referee for carefully reading the manuscript and providing very helpful feedback.
TR and AV acknowledge support from the German Federal Ministry of Economics and Technology (BMWi) provided through the German Space Agency (DLR) under project 50 OR 2112.
AV acknowledges funding by the Deutsche Forschungsgemeinschaft (DFG, German Research Foundation) -- 450861021.
VG acknowledges the financial contribution from the contracts Prin-MUR 2022 supported by Next Generation EU (M4.C2.1.1, n.20227RNLY3 {\it The concordance cosmological model: stress-tests with galaxy clusters}.
MH acknowledges financial support from the Excellence Cluster ORIGINS which is funded by the DFG under Germany’s Excellence Strategy – EXC 2094 – 390783311. AL acknowledges the support from the National Natural Science Foundation of China (Grant No. 12588202). AL is supported by the China Manned Space Program with grant no. CMS-CSST-2025-A04.
We thank F. Maccagni and P. Serra for providing MeerKAT continuum contours of Fornax\,A and MeerKAT HI emission contours of ram pressure stripped galaxies.

This work is based on data from eROSITA, the soft X-ray instrument aboard SRG, a joint Russian-German science mission supported by the Russian Space Agency (Roskosmos), in the interests of the Russian Academy of Sciences represented by its Space Research Institute (IKI), and the Deutsches Zentrum für Luft- und Raumfahrt (DLR). The SRG spacecraft was built by Lavochkin Association (NPOL) and its subcontractors, and is operated by NPOL with support from the Max Planck Institute for Extraterrestrial Physics (MPE).

The development and construction of the eROSITA X-ray instrument was led by MPE, with contributions from the Dr. Karl Remeis Observatory Bamberg \& ECAP (FAU Erlangen-Nuernberg), the University of Hamburg Observatory, the Leibniz Institute for Astrophysics Potsdam (AIP), and the Institute for Astronomy and Astrophysics of the University of Tübingen, with the support of DLR and the Max Planck Society. The Argelander Institute for Astronomy of the University of Bonn and the Ludwig Maximilians Universität Munich also participated in the science preparation for eROSITA.

The eROSITA data shown here were processed using the eSASS software system developed by the German eROSITA consortium.

This research has made use of the NASA/IPAC Extragalactic Database (NED) which is operated by the Jet Propulsion Laboratory, California Institute of Technology, under contract with the National Aeronautics and Space Administration.

Partly based on observations obtained with \textit{XMM-Newton}, an ESA science mission with instruments and contributions directly funded by ESA Member States and NASA.

ChatGPT has been used to speed up writing some of the Python scripts, e.g., for plotting data.

This work made use of astrometry.net.

\end{acknowledgements}

\bibliographystyle{aa}
\bibliography{bibo_engl}
\begin{appendix}
\section{Comparison of eRASS:2 and eRASS3+eRASS4+eRASS5 images}
\label{eRASS:2-345}
\begin{figure*}
    \centering
    \includegraphics[width=\hsize]{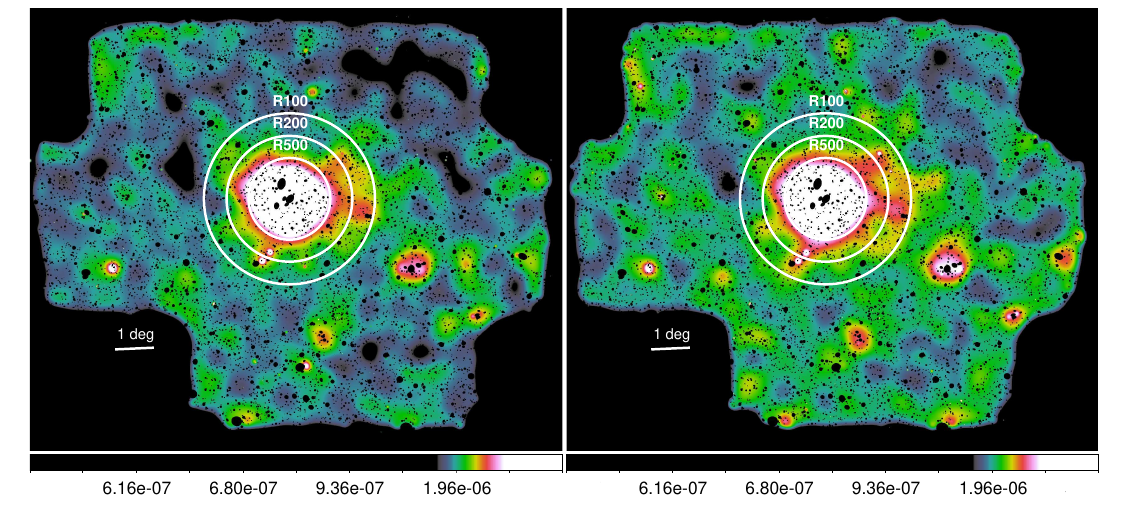}
    \caption{Same as Fig.~\ref{fig:image} (bottom) but for eRASS:2 (left) and eRASS3+eRASS4+eRASS5 (right), i.e., two completely independent datasets.}
    \label{fig:erass345_12}
\end{figure*}

To check qualitatively if some of the features in the eROSITA images are caused by photon noise, we compare in Fig.~\ref{fig:erass345_12} two completely independent images. On the left hand side, we show the image using data from the first two eROSITA all-sky surveys (eRASS:2) and on the right hand side, the one created using data for the sum of the third, fourth, and fifth survey (eRASS3+eRASS4+eRASS5).

Essentially, the same features appear for both images here as those discussed in the main text for the full dataset (eRASS:5). In particular, Fingers 1--3 and Excesses 1--4 are present in both. This demonstrates visually that these surface brightness features are not caused by statistical fluctuations. This is expected given the noise-suppression properties of the wavelet-filtering process.
\FloatBarrier

\section{Hard band eRASS:5 image}
\label{hard_eRASS:5}
\begin{figure}
    \centering
    \includegraphics[width=\hsize]{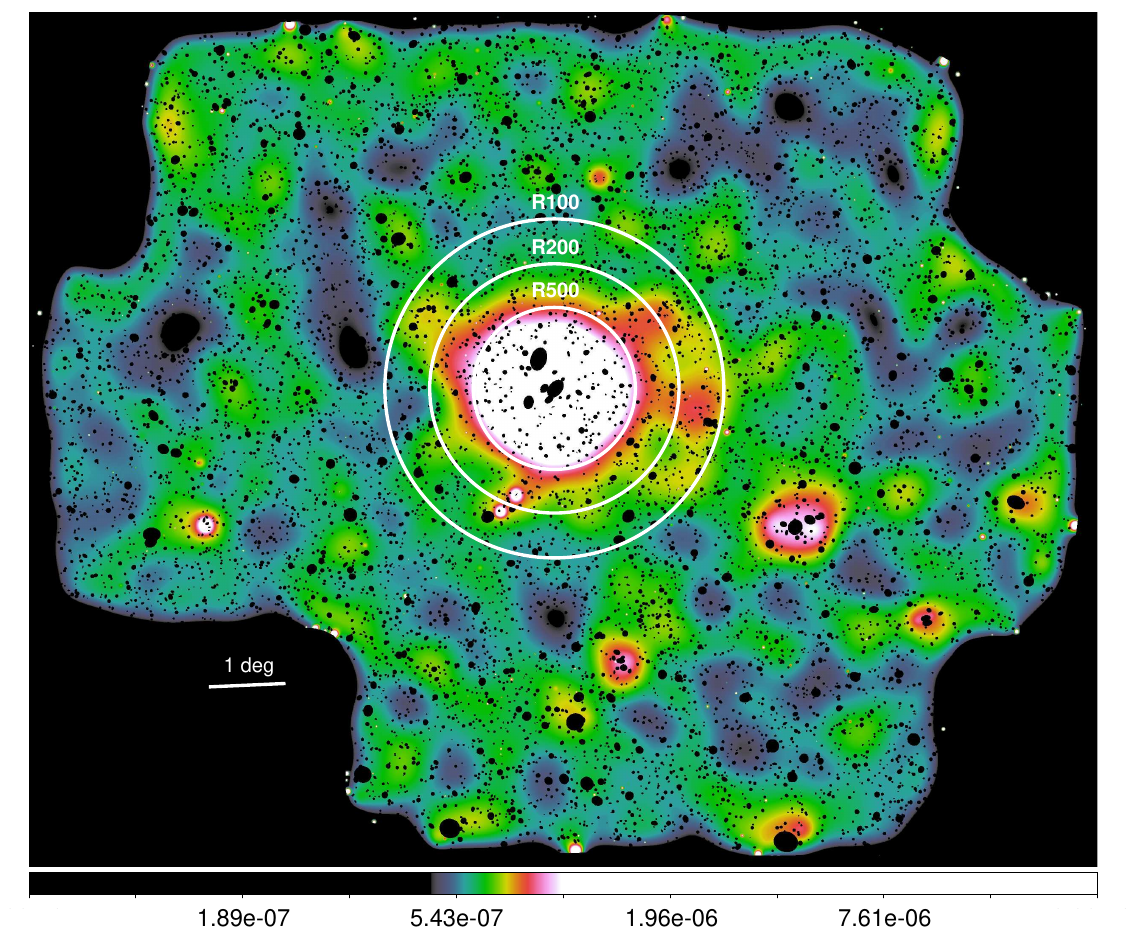}
    \caption{Same as Fig.~\ref{fig:image} (bottom) but for the hard band, 1.0--2.3 keV.}
    \label{fig:hard_eRASS:5}
\end{figure}

In Fig.~\ref{fig:eRASS1_map} we show that the Fornax cluster lies in a region with particularly stable background emission; background is low and does not show obvious strong variations. Still, given the scale of several degrees covered by the eROSITA image of the Fornax cluster, it is difficult to exclude emission variations in our Milky Way Galaxy as the underlying cause of the observed surface brightness fluctuations. Furthermore, varying absorption and uncertainties in our correction procedure may also induce apparent surface brightness variations. Both effects, emission from the low temperature plasma in our Galaxy and absorption by its neutral gas, are in practice only important well below a photon energy of 1 keV. Therefore, if one or both of these effects indeed gave rise to the features discussed in this paper, we would expect them to vanish in an image created using only photons with an energy of at least 1 keV. Therefore, we constructed a hard band image using the energy range 1.0--2.3 keV and show it in Fig.~\ref{fig:hard_eRASS:5}.

Essentially, the same features appear in this image as in the broad band image. All fingers and excess regions are present. Also, the bridge to the Fornax\,A group becomes particularly clear in this energy band.
This demonstrates that these surface brightness features are unlikely to be caused by varying emission or absorption effects in our Galaxy, as these are expected to be very weak above 1.0 keV.
\FloatBarrier

\section{Surface brightness profiles in sectors}
\label{app:sectors}
\begin{figure}
    \centering
    \includegraphics[width=0.8\hsize]{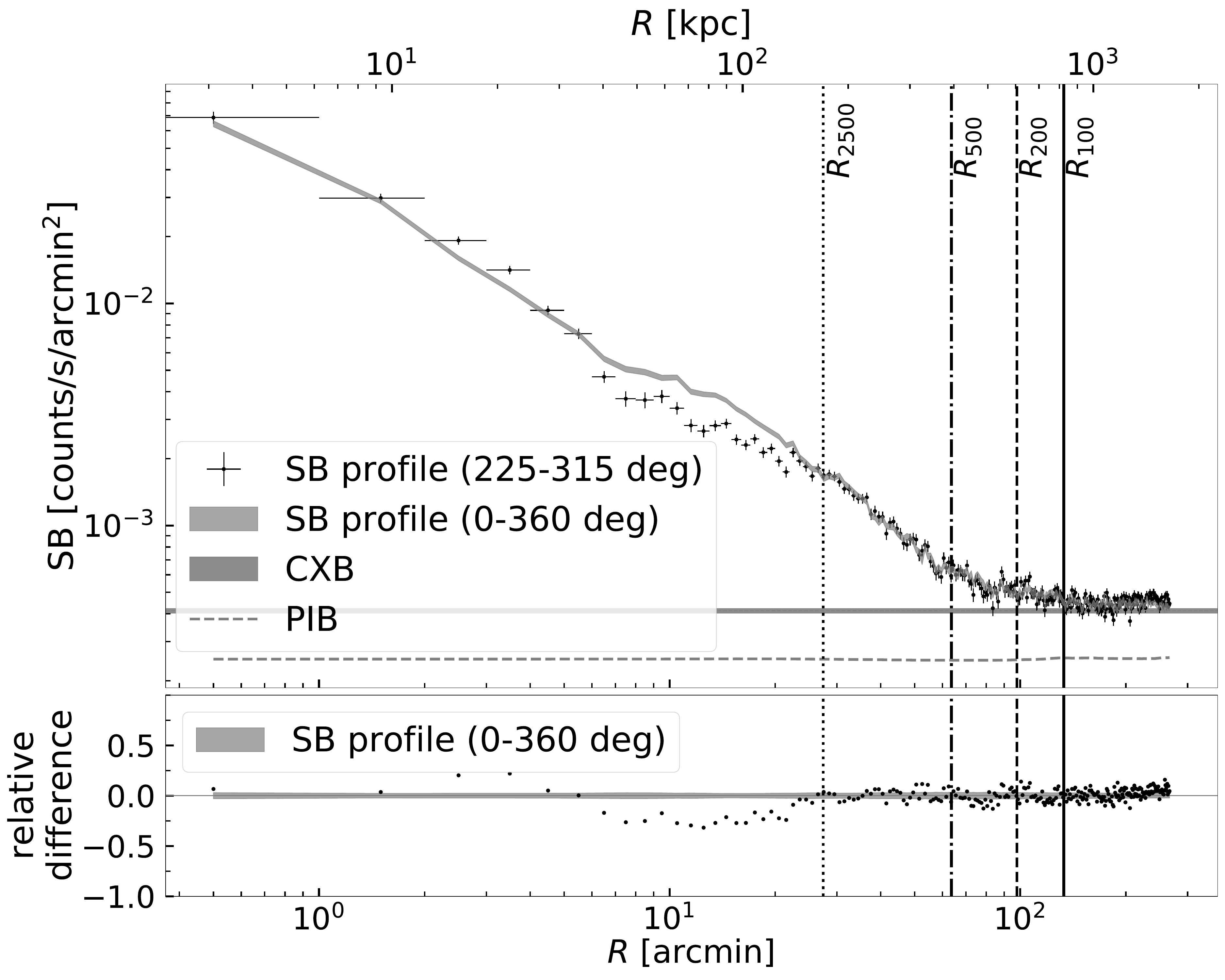}
    \includegraphics[width=0.8\hsize]{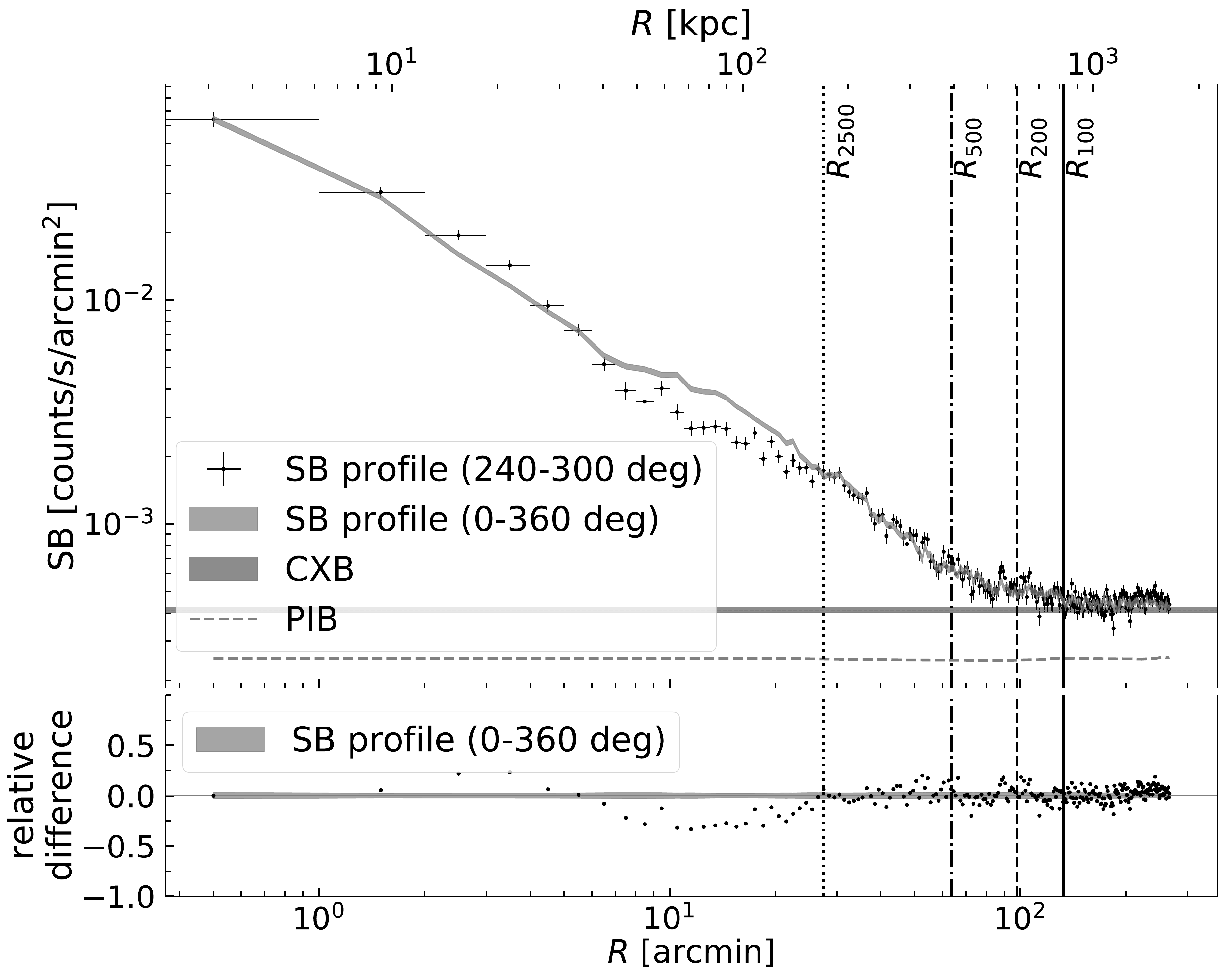}
    \includegraphics[width=0.8\hsize]{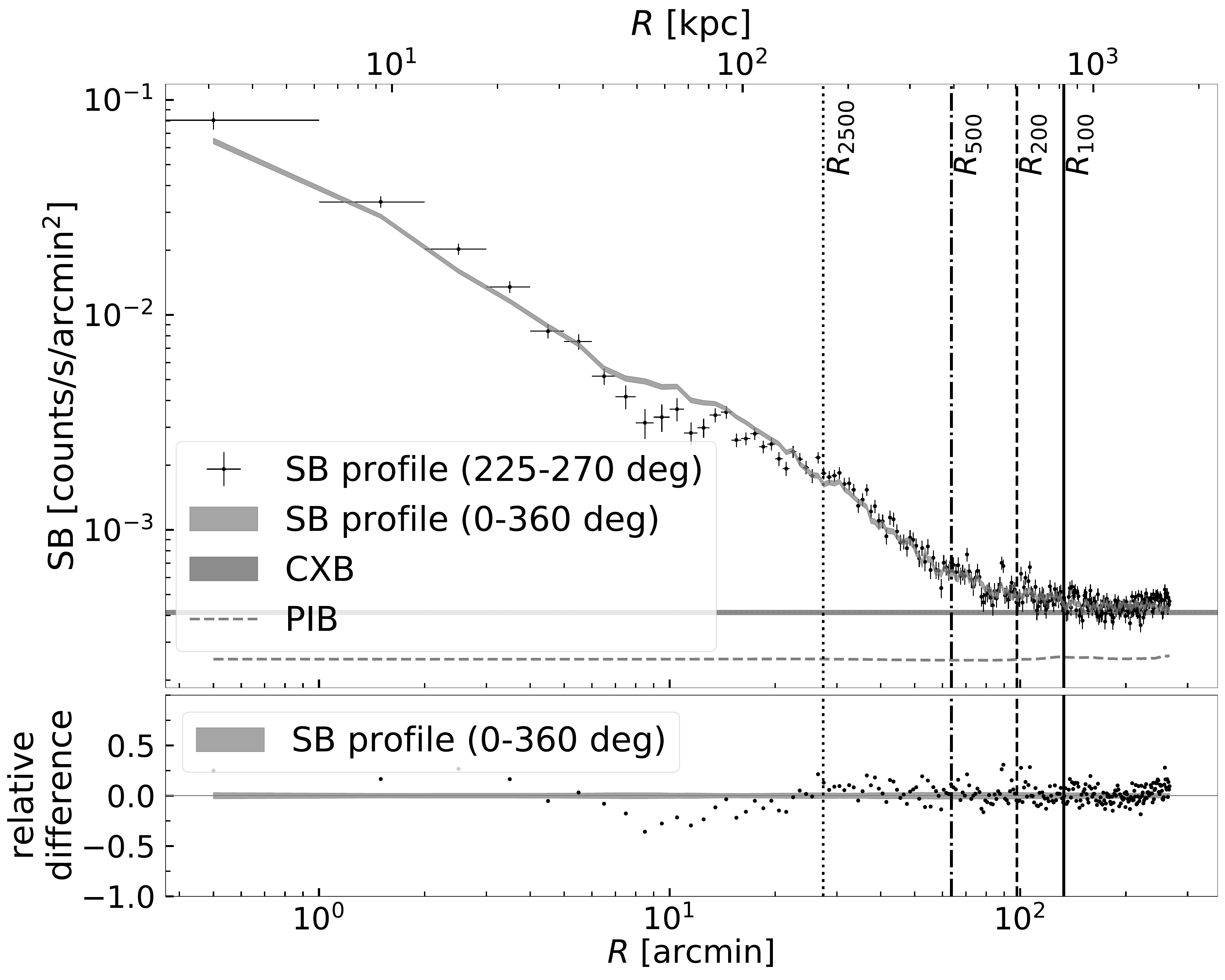}
    \includegraphics[width=0.8\hsize]{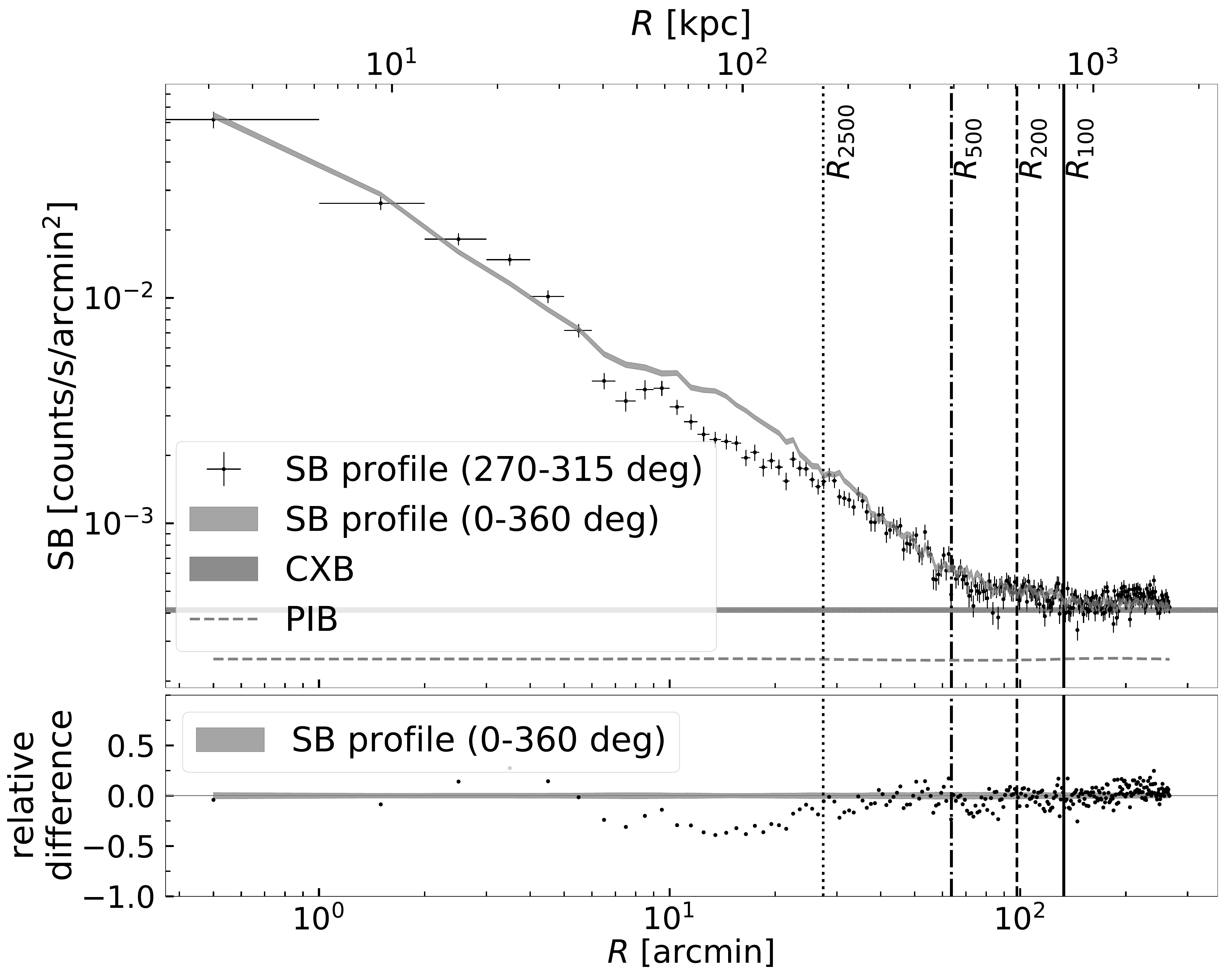}
    \caption{Surface brightness profile sectors in the direction of the predicted bow shock.}
    \label{fig:SB_prof4}
\end{figure}

\begin{figure}
    \centering
    \includegraphics[width=0.8\hsize]{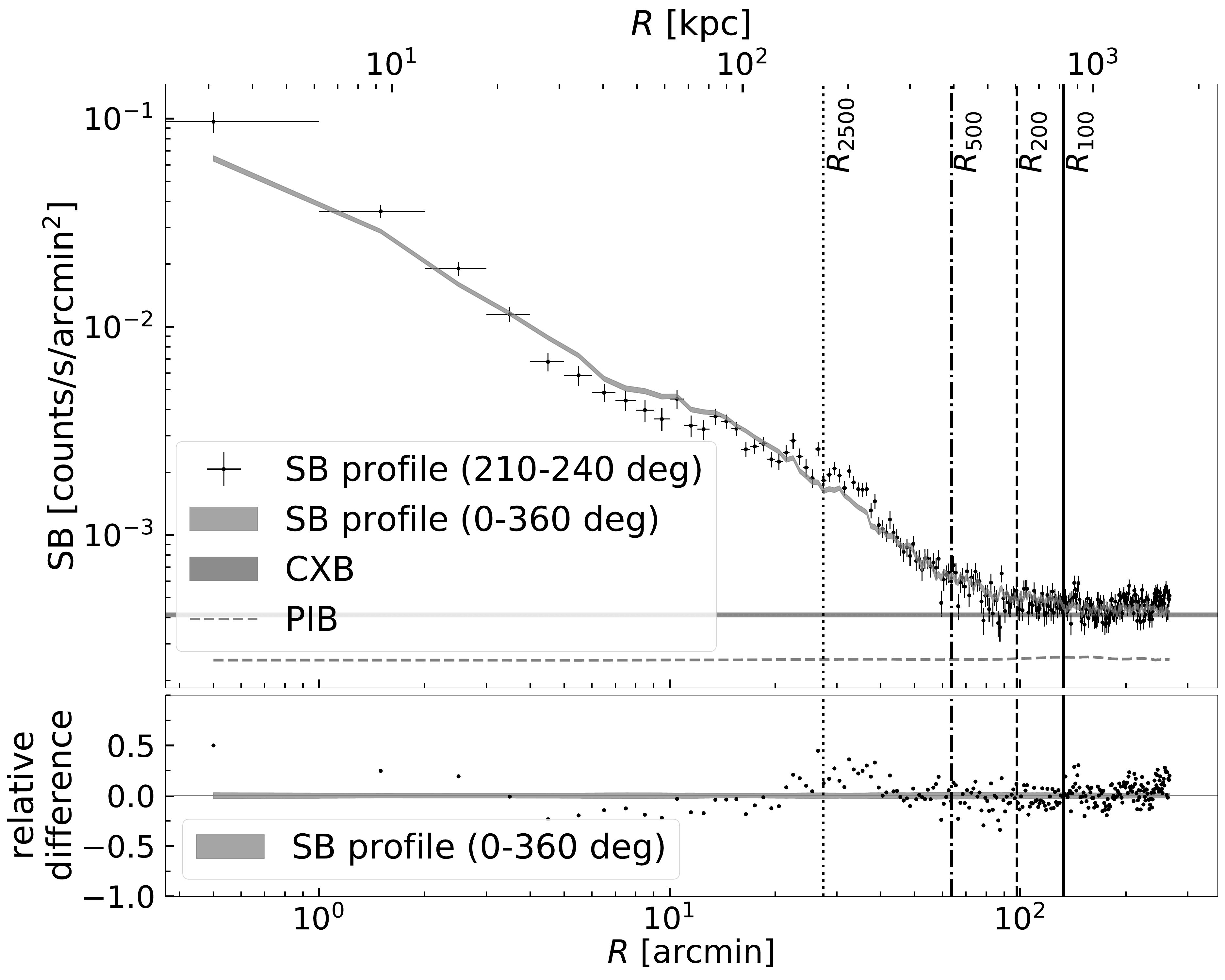}
    \includegraphics[width=0.8\hsize]{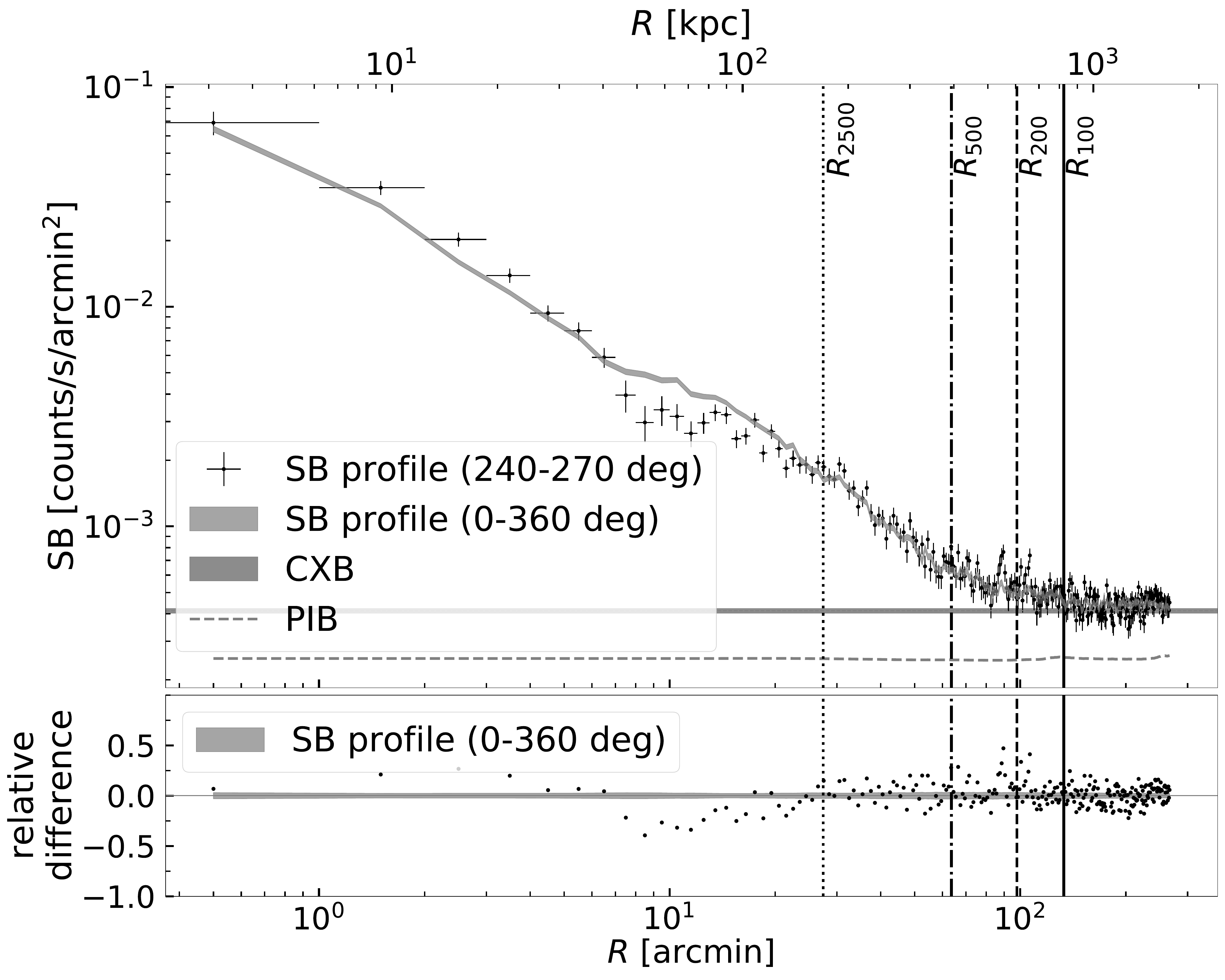}
    \includegraphics[width=0.8\hsize]{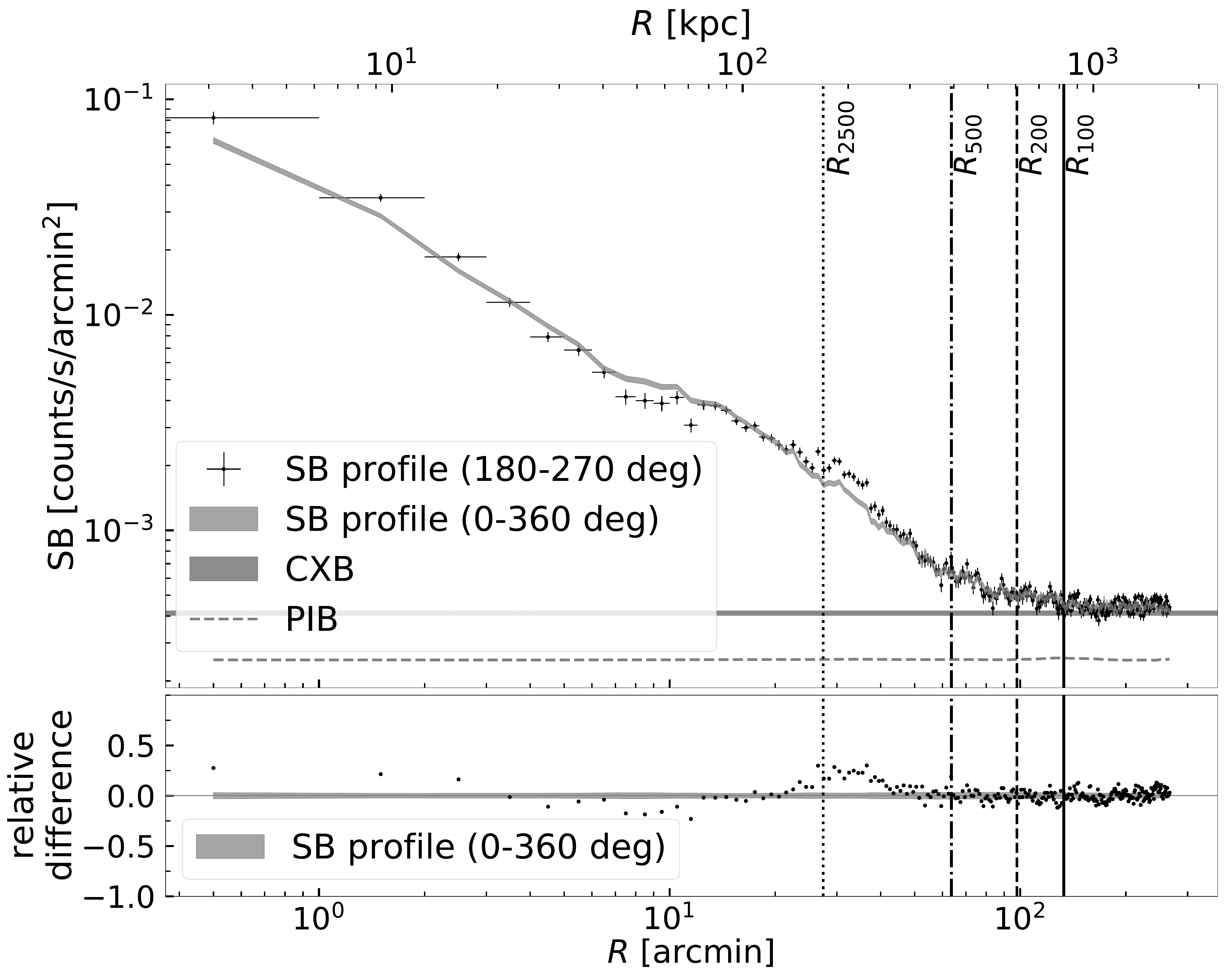}
    \includegraphics[width=0.8\hsize]{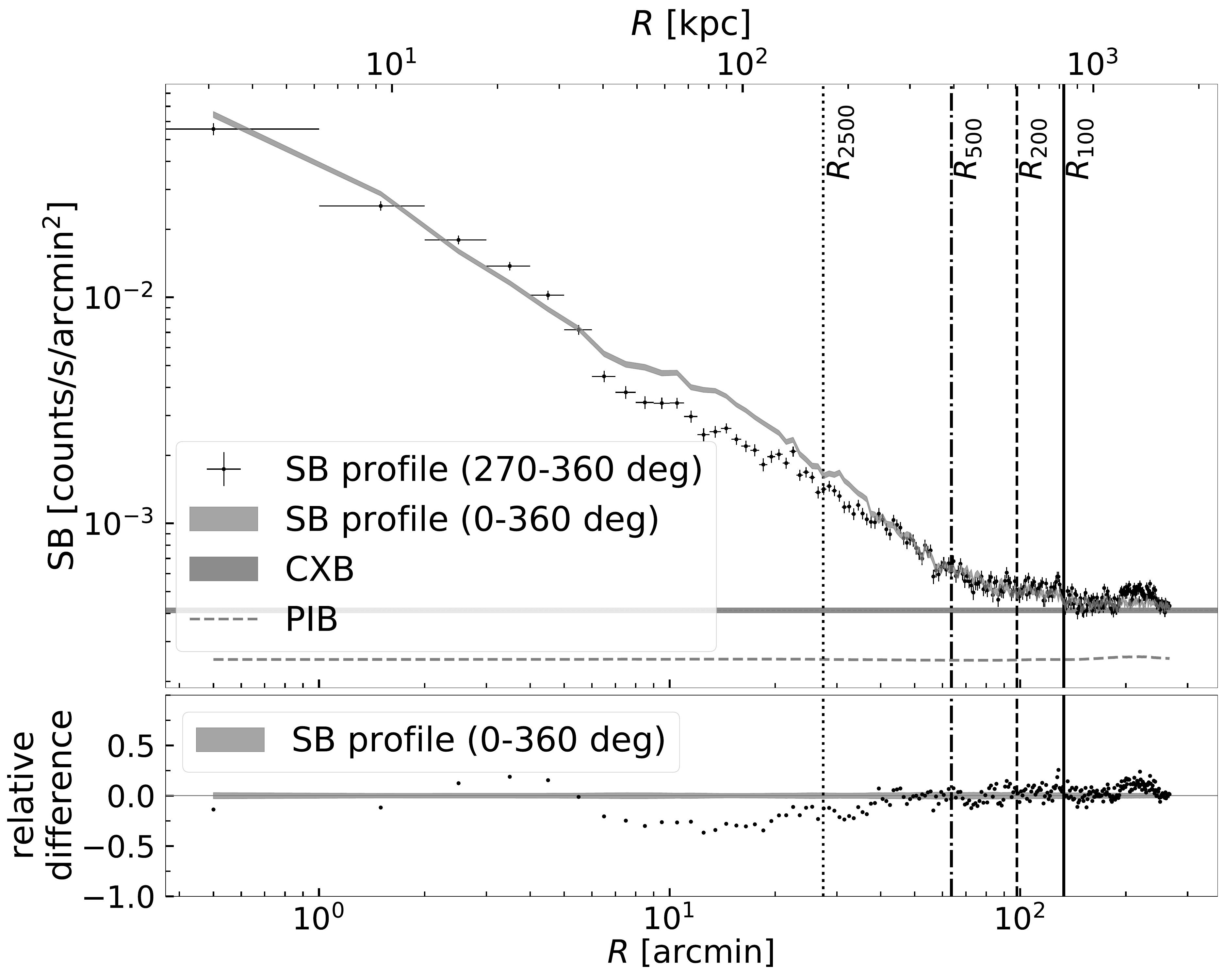}
    \caption{Same as Fig.~\ref{fig:SB_prof4} but showing several more sectors toward the predicted bow shock.}
    \label{fig:SB_prof5}
\end{figure}

As discussed at the end of Sect.~\ref{sec:res:XSB}, we searched for a large-scale bow shock in various directions. We show the surface brightness profiles of several sectors in Figs.~\ref{fig:SB_prof4} and \ref{fig:SB_prof5}.
\FloatBarrier

\section{Close-up views of background clusters}
\label{app:bkg_cl}
eROSITA can detect extended emission from hot gas not only nearby but out to redshifts even beyond one. In Fig.~\ref{fig:erass1} we show the distribution of background clusters on top of the eROSITA image of the Fornax cluster and the comparison was discussed in detail in Sect.~\ref{sec:res:cl}. Here, we provide versions of the same image but zoomed into various areas to provide more detail.
\begin{figure}
    \centering
    \includegraphics[width=\hsize]{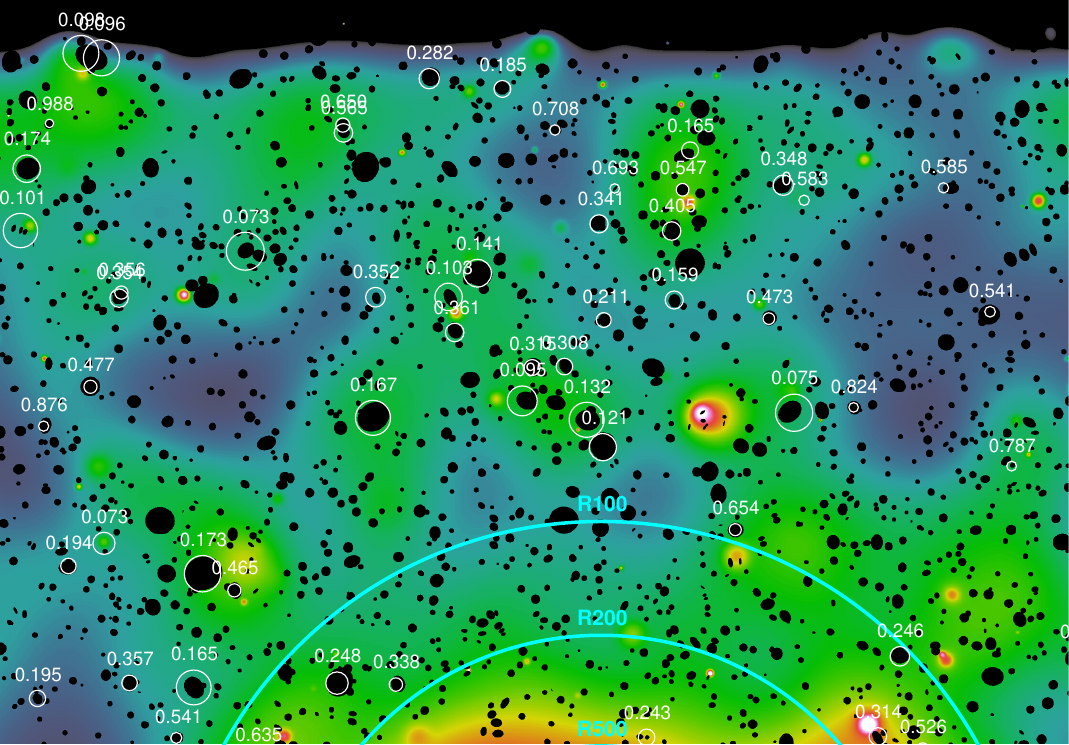}
    \includegraphics[width=\hsize]{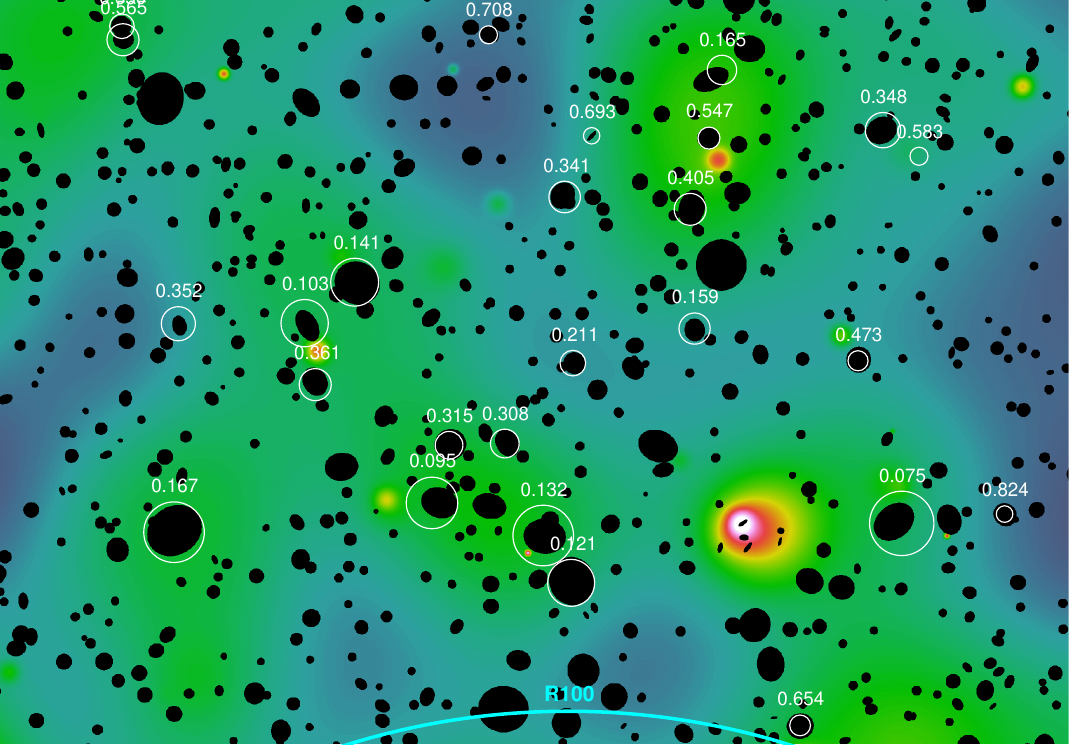}
    \includegraphics[width=\hsize]{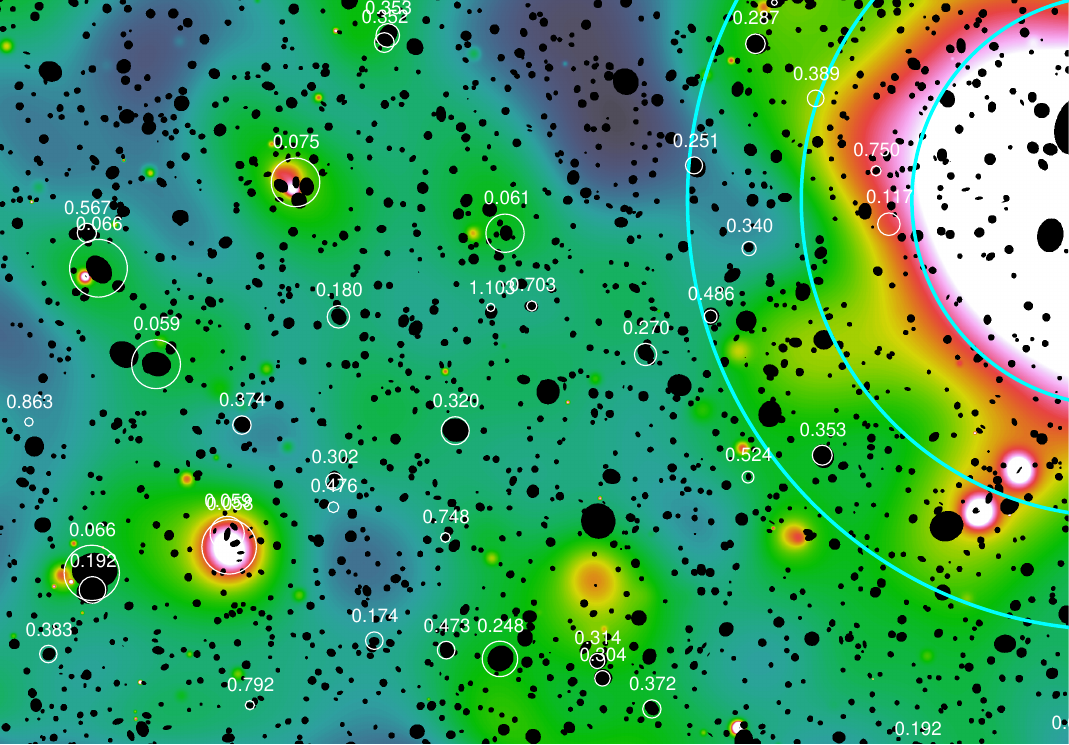}
    \caption{Same as Fig.~\ref{fig:erass1} but zoomed into various areas.
    }
    \label{fig:erass1_zoom1}
\end{figure}
\begin{figure}
    \centering
    \includegraphics[width=\hsize]{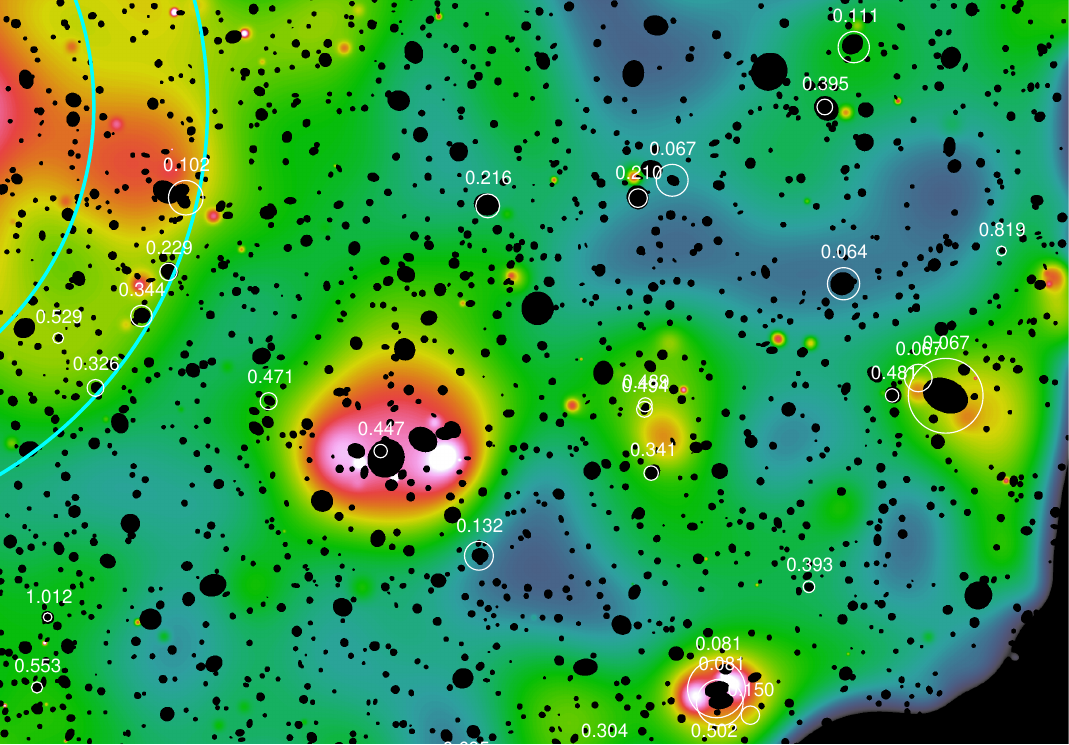}
    \includegraphics[width=\hsize]{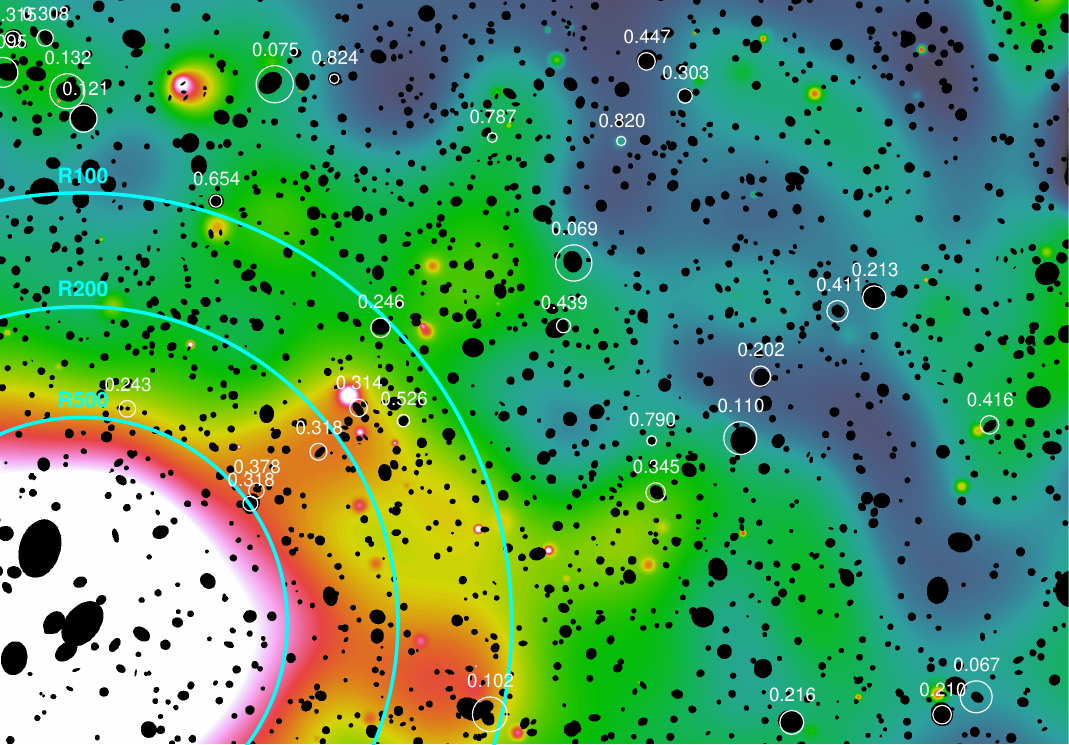}
    \includegraphics[width=\hsize]{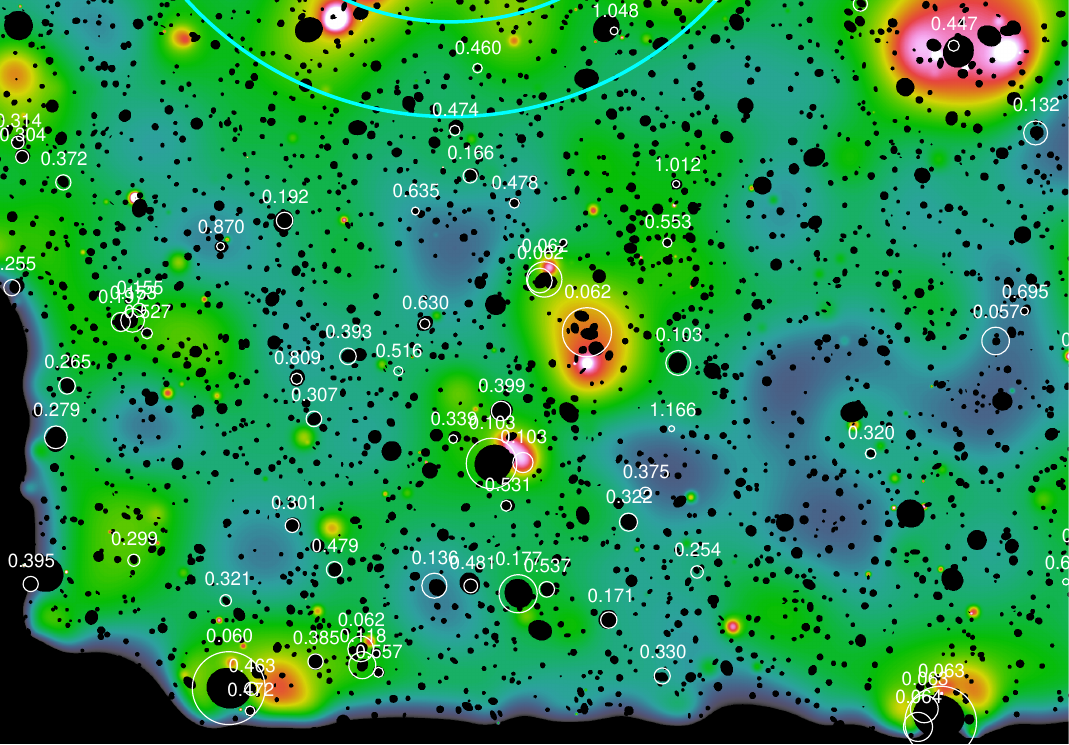}
    \caption{Same as Fig.~\ref{fig:erass1_zoom1} but for a few more areas.}
    \label{fig:erass1_zoom2}
\end{figure}

\end{appendix}

\end{document}